\documentclass[superscriptaddress,amsmath,amssymb,aps,prfluids]{revtex4-2}

\usepackage{algpseudocode}
\usepackage{graphicx}
\usepackage[colorlinks=true, allcolors=blue]{hyperref}
\usepackage{acronym}
\usepackage{tikz}
\usepackage{xcolor}
\usepackage{placeins}

\usepackage{xcolor}

\newcounter{algorithm}
\newcommand{\algorithmcaption}[2]{%
  \refstepcounter{algorithm}%
  \label{#2}%
  \par\vspace{2pt}
  \begin{center}
  \textbf{Algorithm~\thealgorithm.} #1
  \end{center}}

\begin{document}
\noindent\fbox{%
  \parbox{0.6\textwidth}{\small
    This is the accepted manuscript of an article published in:\\
    \textbf{Physical Review Fluids}.\\
    The journal version is available online at:\\
    \href{https://doi.org/10.1103/nbpv-s32c}{\textbf{https://doi.org/10.1103/nbpv-s32c}}\\
    Please cite the journal version rather than the arXiv entry.%
  }%
}
\vspace{1em}

\title{Data-driven impeller model for efficient large eddy simulations\\ of metastable von Kármán flows}

\author{Quentin Malé}
\author{Lucas Amoudruz}
\affiliation{School of Engineering and Applied Sciences, Harvard University, Cambridge, MA 02138, USA}

\author{Daniel Bulgarini}
\affiliation{Department of Aerospace Engineering and Engineering Mechanics, The University of Texas at Austin, Austin, TX 78712, USA}

\author{Shuolin Xiao}
\affiliation{Department of Mechanical Engineering, Johns Hopkins University, Baltimore, MD 21218, USA}

\author{Gregory Eyink}
\affiliation{Department of Applied Mathematics and Statistics, Johns Hopkins University, Baltimore, MD 21218, USA}

\author{Charles Meneveau}
\affiliation{Department of Mechanical Engineering, Johns Hopkins University, Baltimore, MD 21218, USA}

\author{Fabrizio Bisetti}
\affiliation{Department of Aerospace Engineering and Engineering Mechanics, The University of Texas at Austin, Austin, TX 78712, USA}

\author{Petros Koumoutsakos}
\email{petros@seas.harvard.edu}
\affiliation{School of Engineering and Applied Sciences, Harvard University, Cambridge, MA 02138, USA}

\date{\today}

\begin{abstract}
The von Kármán turbulent swirling flow exhibits intriguing large-scale metastable dynamics, including low-frequency state switching. The study of state switching demands long-duration high-fidelity simulations at high Reynolds numbers that capture the flow generated by the impellers. Blade-resolved Large Eddy Simulations (LES) are computationally prohibitive, limiting access to these slow dynamics.
Here, we develop a model for the action of the impellers on the flow using experimental data from Particle Image Velocimetry (PIV) and torque measurements of the von Kármán flow.
The impeller-region velocity is parametrized via B-splines and coupled to the LES through momentum forcing. An initial set of B-spline coefficients is inferred using the Optimizing a DIscrete Loss (ODIL) framework constrained by the Reynolds-Averaged Navier--Stokes (RANS) equations, PIV measurements in the optically accessible portion of the device, and impeller torque measurements. The coefficients are then refined by the Covariance Matrix Adaptation Evolution Strategy (CMA-ES), which minimizes the discrepancy between the LES time-averaged velocity and torque and their experimental counterparts.
Using the data-driven impeller model, we perform long-duration LES of the von Kármán flow. We find that the simulation reproduces the mean flow in the bulk and displays metastable state-switching dynamics. We further show that these metastable states are not axisymmetric and consist of an alternating four-cell flow pattern that slowly rotates around the axis of the cylindrical vessel. The proposed approach provides a practical and computationally efficient route to investigating large-scale dynamics in impeller-driven turbulent flows.
\end{abstract}

\maketitle

\acrodef{DNS}{Direct Numerical Simulation}
\acrodef{LES}{Large Eddy Simulation}
\acrodef{ODIL}{Optimizing a DIscrete Loss}
\acrodef{SGS}{Sub-Grid Scale}
\acrodef{SPIV}{Stereoscopic Particle Image Velocimetry}
\acrodef{PIV}{Particle Image Velocimetry}
\acrodef{RANS}{Reynolds Averaged Navier--Stokes}
\acrodef{PDE}{Partial Differential Equation}
\acrodef{ODE}{Ordinary Differential Equation}
\acrodef{RSE}{Root Square Error}
\acrodef{MSE}{Mean Square Error}
\acrodef{RMSE}{Root Mean Square Error}
\acrodef{RMS}{Root Mean Square}
\acrodef{PDF}{Probability Density Function}
\acrodef{CMA-ES}{Covariance Matrix Adaptation Evolution Strategy}
\acrodef{MAC}{Marker-and-Cell}
\acrodef{EMA}{Exponential Moving Average}

\section{Introduction}

Metastable turbulence is a characteristic feature of many high-dimensional chaotic systems across a wide range of natural and industrial settings, including planetary dynamos~\cite{berhanu2007magnetic,petrelis2009simple,Tobias_2021}, atmospheric flows~\cite{charney1979multiple,majda2006distinct,DetectingNonequilibriumStatesinAtmosphericTurbulence}, 
wakes and airfoil dynamics~\cite{grandemange2013turbulent,rigas2014low,zaman1989natural,busquet2021bifurcation},
and industrial mixing processes~\cite{hasal2000macro,nikiforaki2003origin}. In these systems, the flow can occupy quasi-persistent states for durations that far exceed the characteristic timescales of the underlying turbulent fluctuations. Understanding the mechanisms behind these transitions is critical for predictability, as they often correspond to high-impact geophysical events or shifts in the efficiency of engineering reactors. A prominent example is found in mid-latitude atmospheric circulation, where the flow intermittently switches between distinct weather regimes such as zonal and blocked states~\cite{charney1979multiple,PersistentAnomaliesBlockingandVariationsinAtmosphericPredictability,weeks1997transitions}. These metastable regimes represent a significant challenge in fluid mechanics as well as non-equilibrium statistical mechanics.

The von Kármán turbulent swirling flow, generated between coaxial counter-rotating impellers in a cylindrical vessel, serves as a prototypical laboratory model for these phenomena. Much like atmospheric regimes, the turbulent von Kármán flow can exhibit multistability and spontaneous symmetry breaking, transitioning from a statistically symmetric two-cell state, through asymmetric two-cell configurations with one enlarged recirculation, to fully asymmetric one-cell attractors~\cite{ravelet2004multistability,delatorre2007slow,cortet2010experimental,cortet2011susceptibility}. Numerical simulation presents a unique opportunity to probe the physics of these flows beyond what is currently possible with experimental diagnostics alone. Simulation allows for the study of flow structures in regions that are challenging to access
experimentally, such as the immediate vicinity of the impeller blades, and provides 
spatio-temporal data needed to characterize the complex three-dimensional flow. Capturing the low-frequency, large-scale dynamics of 
state-switching events
is computationally demanding, as it requires the simulation of thousands of impeller revolutions to achieve statistical convergence or to observe the most infrequent 
events. 
Prior numerical simulations of bladed von Kármán flows \cite{kreuzahler_numerical_2014,nore_numerical_2018,kasbaoui_direct_2021,faller_nature_2021,cappanera_turbulence_2021,bousquet2024largescale} have been restricted to tens of impeller revolutions due to the high computational cost of resolving blade geometry, leaving these slow dynamics inaccessible to numerical investigation. 
The computational cost of \ac{DNS} at the relevant Reynolds numbers is presently prohibitive; consequently, \ac{LES} is the only practical option for capturing these dynamics. However, the coarse spatial resolution required to enable long-duration simulations prevents an explicit representation of the von Kármán impeller blade geometry. 
As a result, the action of the impellers must be modeled. A key challenge for this modeling task is our limited knowledge of the fluid dynamics in the impeller region. Experimental diagnostics such as \ac{PIV} provide velocity measurements only in the bulk of the flow. The impeller region, which is the source of 
 angular momentum injection into the flow, is not directly accessible to \ac{PIV} measurement.

To address this challenge, a two-stage data-driven optimization strategy is introduced in this paper. The action of the impellers is represented through a parametrized velocity field defined within a prescribed impeller region, rather than through explicit geometric resolution.
The impeller-region velocity is expressed as a low-order B-spline parametrization and coupled to the \ac{LES} through a volumetric forcing of the momentum equations. The B-spline coefficients are determined in two successive steps.
In the first step, the \ac{ODIL} framework~\cite{karnakov2023modil,karnakov2024odil} is used to infer an initial set of B-spline coefficients by solving an inverse problem constrained 
by the \ac{RANS} equations, experimental velocity measurements in the bulk, and impeller torque measurements. 
In the second step, the B-spline coefficients are refined using the \acf{CMA-ES} coupled directly to the \ac{LES} solver, minimizing the discrepancy between the \ac{LES} time-averaged velocity field and experimental \ac{PIV} data. This \ac{LES}-coupled optimization corrects errors introduced by the \ac{RANS} approximation used in the first step and yields an improved mean-flow prediction.
The \ac{ODIL}-derived coefficients serve two purposes in this second step: i) they initialize the \ac{CMA-ES} search in a physically consistent region of parameter space, and ii) an \ac{LES} run with this forcing produces an established flow field that warm-starts each \ac{CMA-ES} evaluation, reducing the number of impeller revolutions needed to converge time-averaged statistics.
We find that this modeling strategy enables long-duration \ac{LES} of von Kármán flows at high Reynolds number on coarse grids, providing access to slow, large-scale dynamics and metastable state switching that are beyond the reach of blade-resolved simulations at comparable Reynolds numbers and integration times. 
Neither the metastable states nor their transition statistics enter the optimization objective, and no time-dependent signal is imposed to induce transitions. The goal being to investigate the spontaneous emergence and three-dimensional organization of these dynamics.

Furthermore, the present two-stage optimization strategy can be extended to other problems in fluid mechanics where the forcing region is unresolved or experimentally inaccessible and measurements are sparse. In such settings, \ac{ODIL} inference constrained by the \ac{RANS} equations can provide an initialization for a more expensive optimization coupled to \ac{LES}, whose objective function is difficult to differentiate. Examples include compressors, wind turbines, and combustion chambers with swirl injectors. Such an extension concerns the inference and optimization methodology and would require a suitable representation of the forcing region, calibration using device-specific data, and validation for the intended quantities of interest. The present representation does not resolve detailed blade-scale flow phenomena so would only be useful for applications in which such details are not essential.

\section{Methods}

\subsection{von Kármán flow configuration}\label{sec:vk_config}

\begin{figure}[!htbp]
    \centering
    \begin{tikzpicture}
        \node[anchor=south west, inner sep=0] (img)
            {\includegraphics[width=0.85\linewidth]{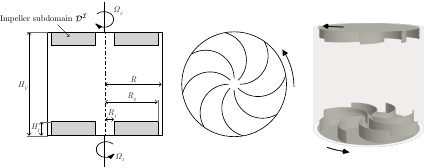}};
        \begin{scope}[x={(img.south east)}, y={(img.north west)}]
            \node[anchor=south west, font=\small] at (0.00, 0) {(a)};
            \node[anchor=south west, font=\small] at (0.4, 0) {(b)};
            \node[anchor=south west, font=\small] at (0.7, 0) {(c)};
        \end{scope}
    \end{tikzpicture}
    \caption{Von Kármán flow configuration. (a) Cylindrical vessel of radius $R$ and height $H_V$ driven by two counter-rotating impellers (angular velocity $\Omega_1$, $\Omega_2$) of height $H_I$. The computational domain $\mathcal{D}$ is partitioned into the impeller subdomain $\mathcal{D}^{\mathcal I}$ and the remaining subdomain $\mathcal{D}^{\mathcal F} := \mathcal{D} \setminus \mathcal{D}^{\mathcal I}$. (b) Bottom view of the curved-blade impeller with inner and outer radii $R_1$ and $R_2$. (c) Schematic three-dimensional view of the device.}
    \label{fig:vktts_diagram}
\end{figure}

The von Kármán flow is established within a closed cylindrical vessel of radius $R$ and height $H_V$, in which the fluid motion is driven by rotating impellers (Fig.~\ref{fig:vktts_diagram}). Two identical impellers of height $H_I$ are mounted coaxially at the bottom and at the top of the vessel. The impeller region extends radially from the inner radius $R_1$ to the outer radius $R_2$.
The impellers act as the sole source of momentum injection into the flow, producing a turbulent swirling motion confined by the cylindrical walls. In the present work, both impellers rotate at the same angular speed and in opposite directions, $\Omega_1 = -\Omega_2 = \Omega$. The Reynolds number is defined based on tip speed as
\begin{equation}
    \mathrm{Re} = \frac{|\Omega| R^2}{\nu} \; \text{,}
\end{equation}
where $\nu$ denotes the kinematic viscosity of the fluid. 
The geometric parameters of the von Kármán flow device used in this work are given in Table~\ref{tab:geometry}.

The method is demonstrated and applied to an operating point at $\mathrm{Re} = 300\,000$, with both impellers spinning with the concave side of the blades leading. This operating condition lies in the fully turbulent regime and exhibits large-scale unsteady dynamics, providing a suitable test case for assessing the ability of the method to reproduce both mean-flow statistics and slow dynamical behavior. The PIV velocity fields used for validation were obtained from the experimental campaign of Cappanera et al.~\cite{cappanera_turbulence_2021}. Further details of the experimental setup are given in Appendix~\ref{appendix:experiments}.

\begin{table}[!htbp]
    \caption{Geometric parameters of the von Kármán flow device used to acquire experimental data \cite{cappanera_turbulence_2021}.}
    \label{tab:geometry}
    \begin{ruledtabular}
    \begin{tabular}{ccccc}
    $R$ & $R_1$ & $R_2$ & $H_V$ & $H_I$ \\
    \colrule
    $100~\mathrm{mm}$ & $10~\mathrm{mm}$ & $92.5~\mathrm{mm}$ & $180~\mathrm{mm}$ & $20~\mathrm{mm}$ \\
    \end{tabular}
    \end{ruledtabular}
\end{table}

\subsection{Large eddy simulation}
\label{sec:method_les}

\acf{LES} is used to model the turbulent flow in the von Kármán configuration. Only the largest scales are resolved on the computational grid, while the effect of unresolved (subgrid/subfilter) scales on the resolved flow is modeled. Given that compressibility effects are negligible at the operating point, the flow is modeled using the incompressible Navier--Stokes equations:
\begin{equation} 
    \nabla \cdot \tilde{\mathbf{u}} = 0 \; \text{,}
    \label{eq:continuity_LES}
\end{equation}
\begin{equation}
    \frac{\partial \tilde{\mathbf{u}}}{\partial t}
    + \nabla \cdot (\tilde{\mathbf{u}}\tilde{\mathbf{u}})
    = -\frac{1}{\rho}\nabla \tilde{p}
    - \nabla \cdot \boldsymbol{\tau}^d
    + \nu \nabla^2 \tilde{\mathbf{u}}
    + \mathbf{f} \; \text{,}
\label{eq:momentum_LES}
\end{equation}
where $\tilde{\cdot}$ denotes a filtered quantity, $\mathbf{u}$ is the velocity vector, $\rho$ is the density, $\tilde{p}=\tilde{p}_*+(1/3)\rho\tau_{kk}$ is the modified pressure where $\tau_{ij}=\widetilde{u_i u_j}-\tilde{u}_i\tilde{u}_j$ is the subgrid-scale tensor, $\tau_{ij}^d=\tau_{ij}-(1/3)\tau_{kk}\delta_{ij}$ is the deviatoric part of the subgrid-scale tensor, $\nu$ is the kinematic viscosity, and $\mathbf{f}$ is a body force. The forcing term is used to steer the velocity in the impeller region via
\begin{equation}
    \mathbf{f} = \alpha (\bar{\mathbf{u}}^{\mathrm{imp}} - \tilde{\mathbf{u}}) \; \text{,}
    \label{eq:forcing_LES}
\end{equation}
where $\bar{\mathbf{u}}^{\mathrm{imp}}$ denotes the target impeller-region velocity,
which remains fixed in time during the LES. The forcing nevertheless responds to the evolving flow through its dependence on the instantaneous resolved velocity $\tilde{\mathbf{u}}$. This response takes the form of a prescribed linear relaxation.
The forcing coefficient is proportional to the magnitude of the impeller angular velocity, $\alpha = C_\alpha |\Omega|$, where $C_\alpha$ is a dimensionless coefficient set to $C_\alpha=1$ in this study. Its influence on the results and the choice of this value are discussed in Appendices~\ref{appendix:alpha_sensitivity} and~\ref{appendix:alpha_optimization}. Although the quantitative statistics of the large-scale instability observed in the vessel depend on $C_\alpha$, the bimodal character of the order-parameter distribution persists over the range of values considered. The present results therefore support a qualitative investigation of the instability, as pursued in the remainder of this work, but do not establish quantitative agreement with its experimental statistics. Determining $C_\alpha$ for such quantitative agreement would require additional experimental characterization of the instability and is left for future work.
The deviatoric part of the subgrid-scale tensor is approximated using the Smagorinsky-Lilly model~\cite{smagorinsky1963general,lilly1967representation}
\begin{equation}
\tau^d_{ij} = -2 (C_S \Delta)^2 |\tilde{\bf S}| \tilde{S}_{ij} \; \text{,}
\end{equation}
where $C_S$ is the Smagorinsky coefficient determined dynamically~\cite{germano1991dynamic} using Lagrangian time averaging \cite{meneveau1996lagrangian}, and $|\tilde{\bf S}|=(2 \tilde{S}_{ij}\tilde{S}_{ij})^{1/2}$ is the magnitude of the filtered strain-rate tensor $\tilde{S}_{ij}$. Walls are modeled using a wall model detailed in Appendix~\ref{appendix:wall_model}, in which the friction velocity $u_\tau$ is determined by solving a one-dimensional boundary-layer equation under the local equilibrium assumption~\cite{wang_moin_2002,cabot_moin_2000,kawai_larsson_2012} but using the appropriately non-dimensionalized formulation and solution fit (the generalized Moody Diagram wall model) of \cite{Meneveau01112020}. 

The governing equations are discretized on a structured staggered grid in cylindrical coordinates using second-order accurate centered finite differences and solved using the massively-parallel code NGA \cite{desjardins_high_2008}. The computational grid is arranged in a cylindrical coordinate system and consists of $N_r \times N_\theta \times N_z = 35 \times 112 \times 62$ cells, where $r$ denotes the radial distance from the axis, $\theta$ is the azimuth, and $z$ is the axial coordinate along the axis.
Assuming statistical stationarity and equating the mean power input per unit fluid mass to the mean dissipation rate yields the global estimate of the Kolmogorov scale $l_\eta = 1.68 \times 10^{-4} R$ for the present operating condition at $\mathrm{Re} = 300\,000$ (Sec.~\ref{sec:vk_config}). The corresponding characteristic grid spacings are $\Delta_r \simeq 165 \eta$, $\Delta_z \simeq 167 \eta$, and, at the outer radius $r=R$, $\Delta_\theta \simeq 333 \eta$.
The grid-sensitivity analysis presented in Appendix~\ref{appendix:grid_sensitivity} supports the use of the $35 \times 112 \times 62$ reference grid: reducing the number of cells in each direction to approximately 75\% leaves the mean flow and large-scale instability statistics nearly unchanged, whereas a reduction to 50\% substantially alters the latter.
Time integration relies on a semi-implicit iterative fractional-step method~\cite{kim_moin_1985}, explicit in the radial and axial directions and implicit in the azimuthal one in order to relax the restrictive stability constraint near the axis. Two sub-iterations per time step are used, yielding nominal second-order accuracy in time for the velocity field. At each sub-iteration, incompressibility is enforced by solving a Poisson equation for the pseudo-pressure and projecting the velocity field onto a divergence-free space.

\subsection{B-spline parametrization of impeller-region velocities}

Within the impeller subdomain $\mathcal{D}^{\mathcal I}$ (Fig.~\ref{fig:vktts_diagram}), the radial and azimuthal velocity components are represented using a low-dimensional B-spline representation in $(r,z)$. This reduces the dimensionality of the parameter space to a small set of coefficients, which parametrize the impeller-region velocity in both the \ac{ODIL} inference and \ac{CMA-ES} optimization steps. Only the radial and azimuthal components of the impeller-region velocity are parametrized; the axial component is determined implicitly through the continuity equation.
Let $\{B_a^{(z)}(z)\}_{a=1}^{n_z}$ and $\{B_b^{(r)}(r)\}_{b=1}^{n_r}$ denote clamped, uniformly spaced B-spline basis functions of degree $p$ defined on the axial interval $|z| \in [H_V/2 - H_I,\, H_V/2]$ and the radial interval $r \in [R_1, R_2]$ spanned by the blades.
For $\eta\in\{r,\theta\}$ we define
\begin{equation}
\bar{u}_{\eta}^{\mathrm{imp}}(r,z)
=
\sum_{a=1}^{n_z}  {\xi}_{\eta,a}\, B_a^{(z)}(z)
+
\sum_{b=1}^{n_r} {\chi}_{\eta,b}\, B_b^{(r)}(r),
\end{equation}
where $\boldsymbol{\xi}_{\eta} \in \mathbb{R}^{n_z}$ and $\boldsymbol{\chi}_{\eta} \in \mathbb{R}^{n_r}$ are vectors of axial and radial spline coefficients, treated as optimization variables. We use quadratic splines ($p=2$) with $n_z=3$ axial basis functions and $n_r=3$ radial basis functions. Because of the symmetry, 
the same axial splines at \(z_{\text{mir}}=-z\) are used with the same coefficients for both regions.
The four coefficient vectors are stacked into a single vector
\begin{equation}
    \mathbf{c} = \bigl(\boldsymbol{\xi}_{r},\; \boldsymbol{\chi}_{r},\; \boldsymbol{\xi}_{\theta},\; \boldsymbol{\chi}_{\theta}\bigr) \in \mathbb{R}^{2(n_z+n_r)} \; \text{,}
    \label{eq:coeff_vector}
\end{equation}
forming the unknown parameters to be determined with \ac{ODIL} and \ac{CMA-ES}.

\subsection{ODIL-based inference}

The \ac{ODIL}-based inference retrieves an effective representation of the mean impeller-region velocities that is consistent with the \ac{RANS} equations and the available experimental data.
Impeller-region velocities $\bar{\mathbf{u}}^\mathrm{imp}$, parametrized through the coefficient vector $\mathbf{c}$ (Eq.~(\ref{eq:coeff_vector})), are inferred by solving an inverse problem. A key challenge is that experimental data from \ac{PIV} provide velocity measurements only in the bulk flow, away from the impeller blades. Thus, $\bar{\mathbf{u}}^\mathrm{imp}$ is retrieved using solutions to the \ac{RANS} equations with the objective that the mean velocity field from \ac{RANS} matches available \ac{PIV} data. The \acf{ODIL} framework~\cite{karnakov2023modil,karnakov2024odil} is used to jointly minimize the discrete \ac{RANS} residual and a data-misfit term quantifying the discrepancy between the \ac{RANS} solution and the experimental measurements.
The resulting set of coefficients $\mathbf{c}^{(0)}$ serves two purposes in the subsequent \ac{CMA-ES} optimization.
First, it initializes the \ac{CMA-ES} search in a physically consistent region of the parameter space.
Second, an \ac{LES} run with this forcing produces an established flow field that is already close to the target state; this flow field is used to warm-start the \ac{LES} evaluations within \ac{CMA-ES}, so that fewer impeller revolutions are required for time-averaged statistics to converge at each candidate evaluation, substantially reducing the computational cost of the optimization.

\paragraph{RANS equations}
The \ac{RANS} equations for a statistically stationary and incompressible flow read as follows
\begin{equation}
    \nabla \cdot \bar{\mathbf{u}} = 0 \; \text{,}
    \label{eq:continuity_RANS}
\end{equation}
\begin{equation}
    \nabla \cdot (\bar{\mathbf{u}}\bar{\mathbf{u}})
    + \frac{1}{\rho}\nabla \tilde{p}
    + \nabla \cdot \boldsymbol{\tau}^{d}_R
    - \nu \nabla^2 \bar{\mathbf{u}}
    = 0 \; \text{,}
    \label{eq:momentum_RANS}
\end{equation}
where $\bar \cdot$ denotes a temporal mean quantity, $\mathbf{u}$ is the velocity, $\rho$ is the density, $ \tilde{p} = \bar p + 2/3 \, \rho k$ is the modified pressure, $\nu$ is the kinematic viscosity and $\boldsymbol{\tau}^{d}_R$ is the deviatoric Reynolds stress tensor which is modeled through an eddy-viscosity closure $\boldsymbol{\tau}^{d}_R = -\nu_t\left(\nabla \bar{\mathbf{u}}+\nabla \bar{\mathbf{u}}^{T}\right)$. The turbulent viscosity $\nu_t$ is modeled using the $k-\varepsilon$ model
\begin{equation}
    \nu_t = C_\mu k^2/\varepsilon \; \text{,}
\end{equation}
\begin{equation}
\nabla \cdot ( k\, \bar{\mathbf{u}})
- \nabla \cdot
\left(
\left(\nu + {\nu_t}/{\sigma_k}\right) \nabla k
\right)
- P_k/ \rho + \varepsilon
=0 \; \text{,}
\label{eq:tke_RANS}
\end{equation}
\begin{equation}
\nabla \cdot ( \varepsilon\, \bar{\mathbf{u}})
- \nabla \cdot
\left(
\left(\nu + {\nu_t}/{\sigma_\varepsilon}\right) \nabla \varepsilon
\right)
- C_1 \varepsilon P_k / \rho k + C_2 \varepsilon^2/k
=0 \; \text{,}
\label{eq:epsilon_RANS}
\end{equation}
where $k$ is the turbulent kinetic energy, $P_k$ its production and $\varepsilon$ its dissipation. The model constants are $C_\mu = 0.09$, $\sigma_k = 1.0$, $\sigma_\varepsilon = 1.3$, $C_1 = 1.44$ and $C_2 = 1.92$~\cite{launder1974numerical}.
Walls are modeled using a friction factor defined as a function of the local Reynolds number~\cite{Meneveau01112020}.

The \ac{RANS} equations are discretized in axisymmetric cylindrical coordinates $(r,z)$ using second-order central finite differences on a staggered \ac{MAC} grid. The computational grid comprises $N_r\times N_z = 70\times 62$ cells in the radial and axial directions, respectively. The radial and axial velocity components $(\bar u_r, \bar u_z)$ are stored on cell faces, while the azimuthal velocity $ \bar u_\theta$, modified pressure $\tilde p$, and turbulence variables $(k,\varepsilon)$ are stored at cell centers. The discretization leads to a nonlinear system of algebraic equations for the discrete unknowns.

\paragraph{Free parameters}

We decompose the computational domain as $\mathcal{D} = \mathcal{D}^{\mathcal I} \cup \mathcal{D}^{\mathcal F}$ with $\mathcal{D}^{\mathcal F} := \mathcal{D} \setminus \mathcal{D}^{\mathcal I}$ where $\mathcal{D}^{\mathcal I}$ denotes the impeller subdomain and $\mathcal{D}^{\mathcal F}$ its complement (Fig.~\ref{fig:vktts_diagram}). The optimization vector comprises both grid-based \ac{RANS} unknowns and the spline coefficients defining the impeller kinematics. The axial velocity component $\bar u_z$ is treated as a grid unknown over the entire domain $\mathcal{D}$, including the impeller region. The radial and azimuthal components are treated differently. In the free-velocity region $\mathcal{D}^{\mathcal F}$, $\bar{u}_r$ and $\bar{u}_\theta$ are standard grid degrees of freedom. In contrast, within the impeller subdomain $\mathcal{D}^{\mathcal I}$ these components are not optimized pointwise. 
Instead, they are prescribed through the spline parametrization $\bar u_r^{\mathrm{imp}}(\cdot;\mathbf{c})$ and $\bar u_\theta^{\mathrm{imp}}(\cdot;\mathbf{c})$, whose coefficient vector $\mathbf{c}$ constitutes a low-dimensional control variable. 
The modified pressure $\tilde p$ and turbulence-model variables $(k, \varepsilon)$ are optimized throughout $\mathcal{D}$.

Accordingly, the optimization variables can be summarized as:
\begin{equation}
\mathbf{v}
=
\Big(
\bar{u}_r|_{\mathcal{D}^{\mathcal F}},\;
\bar{u}_z|_{\mathcal{D}},\;
\bar{u}_\theta|_{\mathcal{D}^{\mathcal F}},\;
\tilde{p}|_{\mathcal{D}},\;
k|_{\mathcal{D}},\;
\varepsilon|_{\mathcal{D}},\;
\mathbf{c}
\Big) \, \text{,}
\end{equation}
where the restrictions indicate the collection of interior \ac{MAC}-grid degrees of freedom in the corresponding subdomain.

\paragraph{Loss functional}

The \ac{ODIL} objective is defined as a weighted least-squares norm of the discrete residuals of the governing equations together with the data-misfit terms. The combination of the two terms can also be formulated in a Bayesian framework as a prior (the equation residuals) and a likelihood term (the data term) \cite{Amoudruz2026Bayesian}. 
Let $\boldsymbol{r}_{\mathrm{m}}$,
$\boldsymbol{r}_{\mathrm{c}}$,
$\boldsymbol{r}_{k}$ and
$\boldsymbol{r}_{\varepsilon}$
denote the discrete momentum, continuity, turbulent kinetic energy
and dissipation residual vectors, respectively.
The continuity equation (Eq.~(\ref{eq:continuity_RANS})) and the $k$--$\varepsilon$ transport equations (Eqs.~(\ref{eq:tke_RANS}) and (\ref{eq:epsilon_RANS})) are enforced throughout the entire domain $\mathcal{D}$, whereas the momentum equations for $(\bar u_r, \bar u_\theta, \bar u_z)$ (Eq.~(\ref{eq:momentum_RANS})) are enforced only in the free-velocity region $\mathcal{D}^{\mathcal F}$.
Inside the impeller region $\mathcal{D}^{\mathcal I}$, the momentum equations are not enforced: $\bar u_r$ and $\bar u_\theta$ are prescribed by the spline parametrization, and $\bar u_z$ adjusts to satisfy continuity.
The resulting loss functional reads
\begin{equation}
\mathcal{J}(\mathbf{v})
=
\frac{1}{N_{\mathrm{m}}}
\left\|
\mathbf{r}_{\mathrm{m}}\big|_{\mathcal{D}^{\mathcal F}}
\right\|_2^2
+
\frac{1}{N_{\mathrm{c}}}
\left\|
\mathbf{r}_{\mathrm{c}}\big|_{\mathcal{D}}
\right\|_2^2
+
\frac{1}{N_{\mathrm{k}}}
\left\|
\mathbf{r}_{k}\big|_{\mathcal{D}}
\right\|_2^2
+
\frac{1}{N_{\mathrm{\varepsilon}}}
\left\|
\mathbf{r}_{\varepsilon}\big|_{\mathcal{D}}
\right\|_2^2
+
w_{\mathrm{PIV}}
\frac{1}{N_{\mathrm{PIV}}}
\left\|
\mathbf{r}_{\mathrm{PIV}}
\right\|_2^2 
+
w_{Q} \, \bigl| r_{Q} \bigr|^2
\, \text{,}
\label{eq:ODIL_residual}
\end{equation}
where $\mathbf{r}_{\mathrm{PIV}}$ and $r_{Q}$ denote the mismatches between the computed solution and the available measurements for the velocity field (PIV) and the impeller torque, respectively. The weights $w_{\mathrm{PIV}}$ and $w_{Q}$ control the relative contributions of the two data-misfit terms to the loss functional.
The data-misfit weights are set to $w_{\mathrm{PIV}}=2$ and $w_{Q}=1 \times 10^{-3}$, 
balancing fidelity to the experimental measurements with enforcement of the governing \ac{PDE} residuals from \ac{RANS}.

\subsection{CMA-ES optimization}

The \ac{ODIL}-inferred set of coefficients $\mathbf{c}^{(0)}$ provides a physically consistent starting point, but it is subject to the modeling error inherent in the \ac{RANS} approximation.
In order to correct this residual error, the B-spline coefficients are refined using the \acf{CMA-ES}~\cite{hansen2016cma}, a derivative-free evolutionary optimization strategy that iteratively adapts a multivariate Gaussian distribution over the parameter space to minimize a black-box objective.
In this refinement, candidate parametrizations are evaluated using \ac{LES} against experimental measurements, so the final calibration does not rely on the \ac{RANS} mean-flow prediction.

The objective function combines the discrepancy between the \ac{LES} time-averaged velocity field and the experimental \ac{PIV} data, the discrepancy between the \ac{LES} time-averaged impeller torque and the experimentally measured torque, and a physics-based penalty on the target impeller-region velocity field,
\begin{equation}
    \mathcal{F}(\mathbf{c}) = \frac{1}{N_p} \left\| \langle\tilde{\mathbf{u}}\rangle_T(\mathbf{c}) - \bar{\mathbf{u}}^e \right\|_2^2
    + w_{Q'} \left( \langle Q \rangle_T(\mathbf{c}) - \bar{Q}^e \right)^2
    + w_\phi\, \phi(\mathbf{c})
    \label{eq:objective_cmaes}
\end{equation}
where $\langle \cdot \rangle_T$ denotes a time average over $T$ impeller revolutions, $\mathbf{c}$ is the B-spline coefficient vector defined in Eq.~(\ref{eq:coeff_vector}), $\langle Q \rangle_T(\mathbf{c})$ is the mean torque transmitted to the fluid by the impeller model and $\bar{Q}^e$ is the corresponding experimentally measured torque. The penalty weight $w_{Q'}$ is set to $w_{Q'}=1\times 10^{-2}$.
Another penalty term $w_\phi\, \phi(\mathbf{c})$ encodes two physical constraints on the B-spline target velocity field $\bar{\mathbf{u}}^\mathrm{imp}(\mathbf{c})$
\begin{gather}
    \phi(\mathbf{c}) = \phi_r(\mathbf{c}) + \phi_\theta(\mathbf{c}) \, \text{with} \\
    \label{eq:physics_penalty}
    \phi_r(\mathbf{c}) = \frac{1}{|\mathcal{D}^\mathcal{I}|}\int_{\mathcal{D}^\mathcal{I}} \max\!\left(0,\,-\bar{u}_r^\mathrm{imp}(\mathbf{c})\right)^2 \mathrm{d}\mathcal{D} \, \text{and} \\
    \phi_\theta(\mathbf{c}) = \frac{1}{|\mathcal{D}^\mathcal{I}|}\int_{\mathcal{D}^\mathcal{I}} \max\!\left(0,\,-\bar{u}_\theta^\mathrm{imp}(\mathbf{c})\,\mathrm{sgn}(\Omega_\mathrm{l})\right)^2 \mathrm{d}\mathcal{D} \, \text{,}
\end{gather}
where $\Omega_\mathrm{l}$ is the angular velocity of the local impeller. The term $\phi_r$ penalizes inward radial velocity, enforcing centrifugal pumping; $\phi_\theta$ penalizes azimuthal velocity opposing the impeller direction of rotation. The penalty weight is set to $w_\phi = 0.1$. 
Without this penalty, fitting the sparse \ac{PIV} data alone can drive the B-spline coefficients toward target velocity fields that locally violate the expected impeller flow physics (e.g., inward radial velocity or rotation-opposing azimuthal velocity), which in turn can destabilize or corrupt the LES near the impeller region. No constraint is imposed on the axial component, as it is not an independent B-spline degree of freedom but is instead recovered from the continuity equation.

At each \ac{CMA-ES} iteration, a population of candidate coefficient vectors is sampled from the current Gaussian distribution, each candidate is evaluated by running the \ac{LES} and computing $\mathcal{F}$, and the distribution mean and covariance matrix are updated to favor regions of lower objective value.
The population size $\lambda$ follows the standard CMA-ES default, $\lambda = 4 + \lfloor 3 \ln n_\text{dim} \rfloor$, which yields $\lambda = 11$ individuals per generation for the $n_\text{dim} = 12$ B-spline parametrization.
The distribution mean is initialized with $\mathbf{c}^{(0)}$ from \ac{ODIL}, so that the search begins from a physically consistent region of the parameter space. The covariance matrix is initialized to the identity, with initial step size $\sigma_{\mathrm{CMAES}}^\mathrm{init} = 0.1$, corresponding to an isotropic initial search distribution.
Each candidate \ac{LES} evaluation is warm-started from the established flow field obtained by running the \ac{LES} with the \ac{ODIL} forcing prior to the \ac{CMA-ES} iterations; since the flow does not need to spin up from rest, fewer impeller revolutions are required to converge the time-averaged velocity field, reducing the cost per function evaluation.
Specifically, each evaluation discards the first $5$ impeller revolutions as a burn-in period and then averages the velocity field over the subsequent $T=35$ revolutions to compute $\mathcal{F}$.
This averaging window was chosen so that the statistical noise in $\mathcal{F}$ arising from finite-time averaging remains substantially smaller than the differences in $\mathcal{F}$ that \ac{CMA-ES} must resolve between competing candidates.
The optimized coefficient vector $\mathbf{c}^\star$ is then used to define the impeller-region forcing for long-duration \ac{LES} of the von Kármán flow.

\section{Results and discussion}

\subsection{ODIL-based inference of impeller-region velocity fields}\label{sec:odil_results}

Prior to solving the inverse problem, a forward \ac{RANS} computation is performed in which a prescribed azimuthal velocity $\bar u_\theta |_{\mathcal{D}^{\mathcal I}} \, (r) = \Omega \, r$ is imposed within the impeller regions ${\mathcal{D}^{\mathcal I}}$. This provides an initial velocity field that serves as the starting point for the subsequent ODIL-based optimization. The inverse procedure then relaxes this constraint and infers the effective velocity distribution that best satisfies both the governing equations and the experimental data.

\begin{figure}[!htbp]
    \centering
    \begin{tikzpicture}
        \node[anchor=south west, inner sep=0] (img)
            {\includegraphics[width=0.85\linewidth]{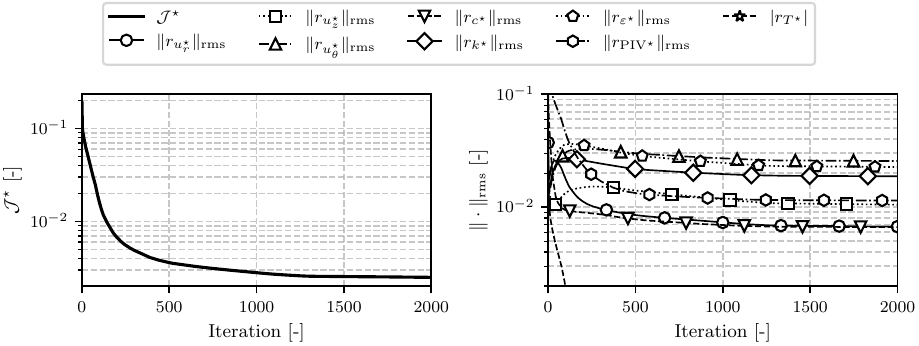}};
        \node[anchor=south west, font=\small, xshift=3mm] at (img.south west) {(a)};
        \node[anchor=south west, font=\small, xshift=3mm] at (img.south) {(b)};
    \end{tikzpicture}
\caption{
Convergence history of the RANS-ODIL flow reconstruction.
(a) Evolution of the objective function constructed from nondimensional residuals $\mathcal{J}^\star$.
(b) Evolution of the \ac{RMS} of the residual components (Eq.~(\ref{eq:ODIL_residual})) with $\| \cdot \|_{\text{RMS}} = (N^{-1}\sum_i^N (\cdot)_i^2)^{1/2}$.
The monotonic decay of the objective and stabilization of the residual norms indicate convergence of the coupled RANS-data optimization.
}
    \label{fig:odil_residuals}
\end{figure}

\begin{figure}[!htbp]
    \centering
    \begin{tikzpicture}
        \node[anchor=south west, inner sep=0] (img)
            {\includegraphics[width=0.75\linewidth]{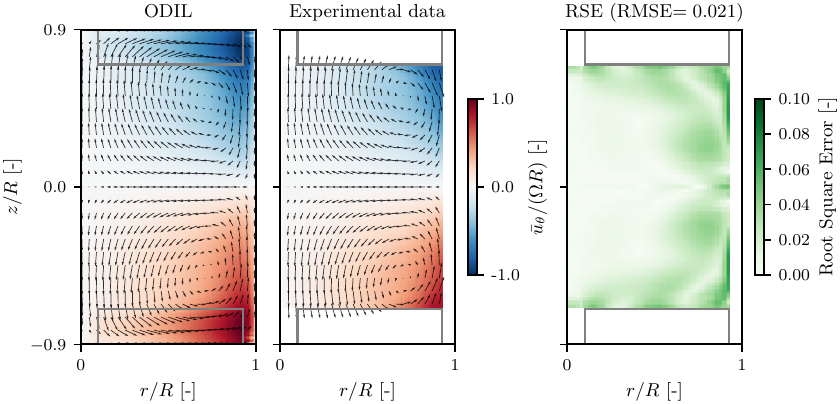}};
        \begin{scope}[x={(img.south east)}, y={(img.north west)}]
            \node[anchor=south west, font=\small] at (0.00, 0) {(a)};
            \node[anchor=south west, font=\small] at (0.31, 0) {(b)};
            \node[anchor=south west, font=\small] at (0.645, 0) {(c)};
        \end{scope}
    \end{tikzpicture}
    \caption{Comparison of the mean velocity field in the meridional plane. (a) \ac{ODIL} reconstruction, (b) experimental measurements, and (c) \acf{RSE} between \ac{ODIL} and experiments. Fields are shown in the ($r/R$, $z/R$) coordinates, with in-plane velocity vectors superimposed on the \ac{ODIL} and experimental mean azimuthal velocity $\bar{u}_{\theta}/(\Omega R)$.}
    \label{fig:odil_velocity_ut_quiver}
\end{figure}

The nonlinear least-squares problem is solved using the Gauss-Newton algorithm described in Appendix~\ref{appendix:gn_odil}. Convergence is achieved after approximately $1\,600$ iterations, as evidenced by the monotonic decay and subsequent stabilization of the objective functional and of the individual residual norms (Fig.~\ref{fig:odil_residuals}). No residual stagnation or oscillatory behavior is observed. The ODIL-based optimization then yields an effective velocity field within the prescribed impeller regions that represents the net action of the rotating blades on the flow (Fig.~\ref{fig:odil_velocity_ut_quiver}). 
The spatial distribution of the error remains weak throughout the bulk, with slightly larger deviations localized near the outer radial boundary and in the vicinity of the impeller edges, where strong velocity gradients are present. 
The inferred velocity components exhibit spatial distributions that are consistent with the expected impeller-induced motion. 
The azimuthal component reflects the primary role of the blades in injecting angular momentum into the flow, increasing with radius and reaching a maximum near the outer impeller radius, where blade-tip effects are strongest.
The radial component contributes to the cross-stream transport needed to sustain axial pumping and the global recirculation. The axial velocity component, retrieved from the continuity constraint, represents the axial pumping implied by the inferred cross-stream motion.
The \ac{ODIL} solution provides a physically consistent starting point for the \ac{CMA-ES} optimization, which corrects residual errors introduced by the \ac{RANS} approximation.

\subsection{LES with ODIL-inferred impeller-region velocity parametrization}

\begin{figure}[!htbp]
    \centering
    \begin{tikzpicture}
        \node[anchor=south west, inner sep=0] (img)
            {\includegraphics[width=0.75\linewidth]{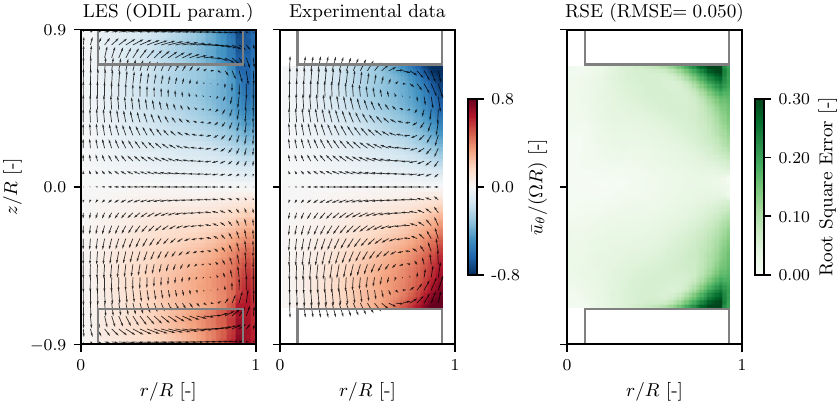}};
        \begin{scope}[x={(img.south east)}, y={(img.north west)}]
            \node[anchor=south west, font=\small] at (0.00, 0) {(a)};
            \node[anchor=south west, font=\small] at (0.31, 0) {(b)};
            \node[anchor=south west, font=\small] at (0.645, 0) {(c)};
        \end{scope}
    \end{tikzpicture}
    \caption{Comparison of the mean velocity field in the meridional plane. (a) \ac{LES} with ODIL-inferred impeller-region velocity parametrization $\mathbf{c}^{(0)}$, (b) experimental measurements, and (c) \acf{RSE} between \ac{LES} and experiments. Fields are shown in the ($r/R$, $z/R$) coordinates, with in-plane velocity vectors superimposed on the \ac{LES} and experimental mean azimuthal velocity $\bar{u}_{\theta}/(\Omega R)$.}
    \label{fig:les_odil_velocity_ut_quiver}
\end{figure}

The ODIL-inferred coefficient vector $\mathbf{c}^{(0)}$ (Eq.~(\ref{eq:coeff_vector})) is used to define the impeller forcing in \ac{LES} of the von Kármán flow. The mean velocity field obtained from this \ac{LES} is compared against the experimental \ac{PIV} data (Fig.~\ref{fig:les_odil_velocity_ut_quiver}).
The \ac{LES} reproduces the large-scale structure of the flow: the two counter-rotating recirculation cells are present, the azimuthal velocity magnitude and its spatial distribution are consistent with the measurements, and the meridional circulation correctly connects the impeller regions to the bulk.
The global \ac{RMSE} between the \ac{LES} and the experimental data is $0.05\,\Omega R$, and the \ac{RSE} map shows that errors are largest near the inner axial edge of the impeller and the outer radial boundary, where strong velocity gradients are present, while the bulk of the domain is well captured.
This result demonstrates that the B-spline parametrization initialized through a \ac{RANS}-based inverse problem solved with \ac{ODIL} already provides a physically consistent and quantitatively accurate representation of the impeller forcing. The \ac{CMA-ES} optimization is subsequently performed to correct the modeling error introduced by the \ac{RANS} approximation used in the \ac{ODIL}-based inference.

\subsection{CMA-ES optimization}\label{sec:cmaes_results}

\begin{figure}[!htbp]
    \centering
    \includegraphics[width=0.45\linewidth]{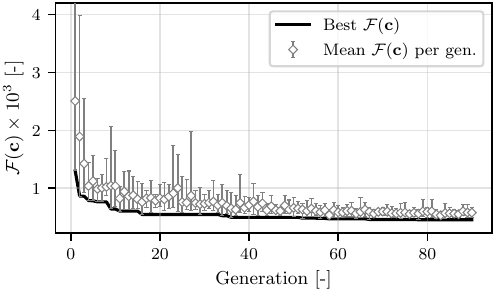}
    \caption{Evolution of the CMA-ES objective function $\mathcal{F}(\mathbf{c})$ (Eq.~(\ref{eq:objective_cmaes})) over optimization generations. The solid line shows the best objective value found up to each generation (overall best). Diamond markers show the mean $\mathcal{F}(\mathbf{c})$ within each generation, with error bars spanning the minimum and maximum values.
    }
    \label{fig:cmaes_quality_vs_batch}
\end{figure}

Starting from the ODIL-inferred coefficient vector $\mathbf{c}^{(0)}$ and the established flow field from the corresponding \ac{LES}, the \ac{CMA-ES} optimization is run to refine the B-spline parametrization.
The population mean of the objective function $\mathcal{F}(\mathbf{c})$ (Eq.~(\ref{eq:objective_cmaes})) decreases rapidly over the first generations and stabilizes after approximately $60$ generations (Fig.~\ref{fig:cmaes_quality_vs_batch}), indicating convergence of the optimization.
One evaluation spans $40$ impeller revolutions, corresponding to approximately $0.75$ hours of wall time on $56$ cores on Intel Xeon Platinum 8280 processors ($41.94$ core-hours). With $11$ candidates per generation the \ac{LES} cost is approximately $461$ core-hours per \ac{CMA-ES} generation.

\begin{figure}[!htbp]
    \centering
    \begin{tikzpicture}
        \node[anchor=south west, inner sep=0] (img)
            {\includegraphics[width=0.75\linewidth]{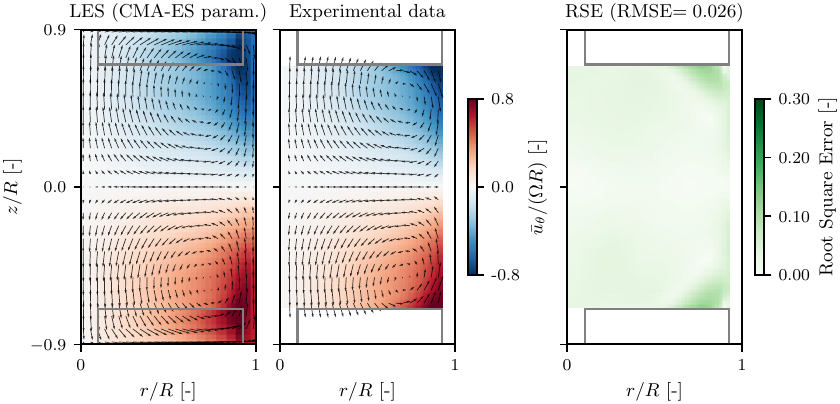}};
        \begin{scope}[x={(img.south east)}, y={(img.north west)}]
            \node[anchor=south west, font=\small] at (0.00, 0) {(a)};
            \node[anchor=south west, font=\small] at (0.31, 0) {(b)};
            \node[anchor=south west, font=\small] at (0.645, 0) {(c)};
        \end{scope}
    \end{tikzpicture}
    \caption{Comparison of the mean velocity field in the meridional plane. (a) \ac{LES} with \ac{CMA-ES} optimized impeller-region velocity parametrization $\mathbf{c}^\star$, (b) experimental measurements, and (c) \acf{RSE} between \ac{LES} and experiments. Fields are shown in the ($r/R$, $z/R$) coordinates, with in-plane velocity vectors superimposed on the \ac{LES} and experimental mean azimuthal velocity $\bar{u}_{\theta}/(\Omega R)$.}
    \label{fig:les_cmaes_velocity_ut_quiver}
\end{figure}

The mean velocity field obtained from the \ac{LES} with the \ac{CMA-ES} optimized coefficients $\mathbf{c}^\star$ is compared against the experimental \ac{PIV} data (Fig.~\ref{fig:les_cmaes_velocity_ut_quiver}). The \ac{RMSE} is reduced from $0.050\,\Omega R$ (ODIL) to $0.027\,\Omega R$ (CMA-ES), a $46$\% improvement. 
Note that these \ac{LES} errors cannot be directly compared with that of the \ac{ODIL} reconstruction (Fig.~\ref{fig:odil_velocity_ut_quiver}(c)), where the entire flow field was constrained by the data-misfit term as part of the flow reconstruction in the impeller region. Here, the errors are obtained from forward simulations governed by the \ac{LES} equations only (Eqs.~(\ref{eq:continuity_LES}) and (\ref{eq:momentum_LES})), and are the relevant measure of accuracy: the forward \ac{LES} is the actual model being assessed. 
The two counter-rotating recirculation cells are well reproduced, the azimuthal velocity and its radial variation are in close agreement with the measurements, and the meridional velocity vectors show the correct large-scale circulation topology throughout the bulk.

\begin{figure}[!htbp]
    \centering
    \begin{tikzpicture}
        \node[anchor=south west, inner sep=0] (img)
            {\includegraphics[width=0.325\linewidth]{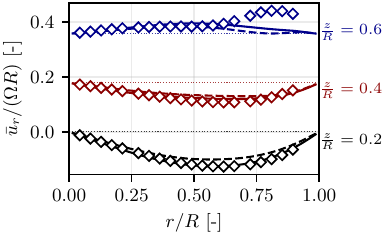}};
        \node[anchor=south west, font=\small] at (img.south west) {(a)};
    \end{tikzpicture}
    \hfill
    \begin{tikzpicture}
        \node[anchor=south west, inner sep=0] (img)
            {\includegraphics[width=0.325\linewidth]{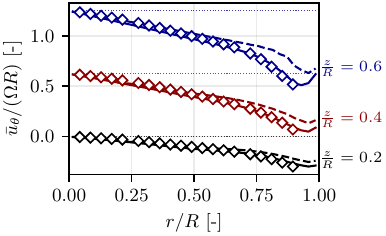}};
        \node[anchor=south west, font=\small] at (img.south west) {(b)};
    \end{tikzpicture}
    \hfill
    \begin{tikzpicture}
        \node[anchor=south west, inner sep=0] (img)
            {\includegraphics[width=0.325\linewidth]{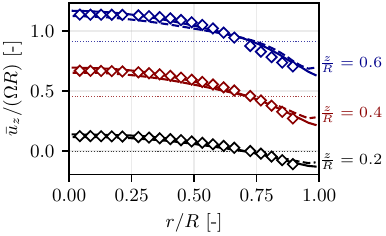}};
        \node[anchor=south west, font=\small] at (img.south west) {(c)};
    \end{tikzpicture}
    \caption{Radial profiles of the mean velocity components (a) $\bar{u}_r/(\Omega R)$, (b) $\bar{u}_\theta/(\Omega R)$, and (c) $\bar{u}_z/(\Omega R)$ at three axial positions $z/R$. Profiles are offset by the local zero-velocity baseline for clarity (cascade representation). Symbols: PIV experiments; dash-dotted lines: LES with ODIL-inferred impeller-region velocity parametrization $\mathbf{c}^{(0)}$; solid lines: LES with CMA-ES-optimized impeller-region velocity parametrization $\mathbf{c}^\star$.}
    \label{fig:radial_profiles}
\end{figure}

The improvement is further confirmed by the radial profiles of all three mean velocity components at three axial positions (Fig.~\ref{fig:radial_profiles}). The LES solution with CMA-ES-optimized parametrization $\mathbf{c}^\star$ consistently lies closer to the experimental data than the baseline LES solution with ODIL-inferred parametrization $\mathbf{c}^{(0)}$ across all components and axial positions. The azimuthal velocity profiles show the largest improvement, with the optimized solution accurately capturing the radial variation toward the outer wall. The radial and axial velocity profiles show modest improvement. Remaining discrepancies are localized near the outer radial boundary and the impeller edges, where the steep velocity gradients are expected to be most sensitive to the absence of explicit blade geometry. 
Specifically, at $z/R=0.6$, both simulations underestimate the radial-velocity peak near the outer wall. This persistent discrepancy may reflect limitations of the reduced impeller representation. 
Nevertheless, the accuracy achieved in the bulk is sufficient to sustain the correct large-scale circulation patterns and to enable the emergence of intrinsic low-frequency dynamics, which are the objective of the present modeling work and are examined in the following section.

\subsection{Observations of metastable flow states and large-scale transitions}

\begin{figure}[!htbp]
    \centering
    \begin{tikzpicture}
        \node[anchor=south west, inner sep=0] (img)
            {\includegraphics[width=0.48\linewidth]{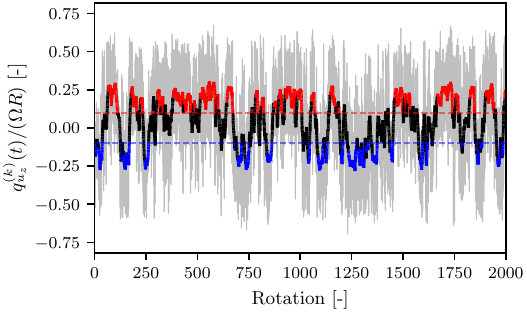}};
        \node[anchor=south west, font=\small] at (img.south west) {(a)};
    \end{tikzpicture}
    \hfill
    \begin{tikzpicture}
        \node[anchor=south west, inner sep=0] (img)
            {\includegraphics[width=0.48\linewidth]{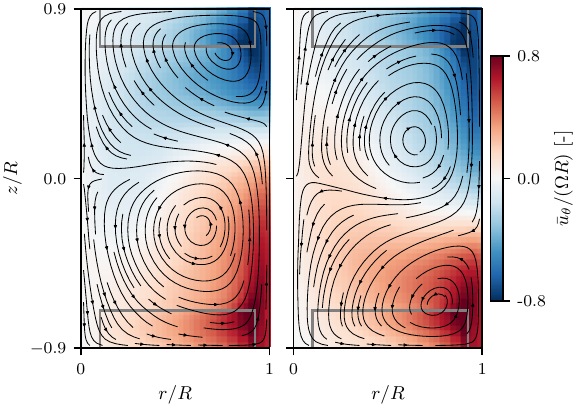}};
        \node[anchor=south west, font=\small] at (img.south west) {(b)};
    \end{tikzpicture}
    \caption{Time evolution and representative flow states of the von Kármán flow at angular position $\theta_k=\pi/4$.
    (a) Time series of the dimensionless order parameter $q_{u_z}^{(k)}(t)/\Omega R$ (gray) together with its exponential moving average colored by the state of the system (red: state A, black: undefined, blue: state B). (b) Streamlines of the mean meridional velocity field $(u_r,u_z)$ conditioned on the state of the system (A/B), illustrating two metastable flow states. In both cases, the flow consists of two recirculation cells of unequal strength, with the dominant circulation located either above or below the mid-plane.}
    \label{fig:order_parameter_signal_and_velocity_conditional_mean}
\end{figure}

\begin{figure}[!htbp]
    \centering
    \includegraphics[width=0.45\linewidth]{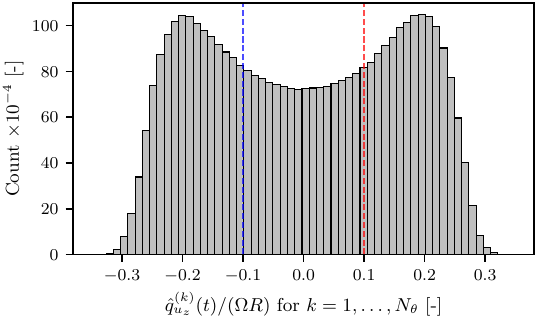}
    \caption{\ac{PDF} of the \ac{EMA}-filtered dimensionless order parameter $\hat{q}_{u_z}^{(k)}(t)/\Omega R$ computed from long-duration LES ($4,500$ rotations) and assembled by sampling at all $k=1,\ldots,N_\theta$. The bimodal distribution reflects the coexistence of two symmetry-related metastable states with opposite dominance of the large-scale circulation.}
    \label{fig:order_param_histogram}
\end{figure}

We now study the dynamics of the flow obtained from the LES with the \ac{CMA-ES} optimized coefficient vector $\mathbf{c}^\star$.
The simulation spans $4,500$ impeller revolutions, corresponding to approximately $84$ hours of wall time on $56$ cores on Intel Xeon Platinum 8280 processors ($4,720$ core-hours).
Over this long duration, we observe metastable large-scale flow states and intermittent transitions between them (Fig.~\ref{fig:order_parameter_signal_and_velocity_conditional_mean}(a)).
To characterize these metastable states, an order parameter is constructed from the azimuthally and axially averaged axial velocity sampled in the near-wall annular region $0.90 \le r/R \le 0.95$, $|z| < z_\mathrm{lim}$, where $z_\mathrm{lim}/R = 0.5$ excludes the impeller forcing regions. For each azimuthal plane at angular position $\theta_k$ ($k = 1,\ldots,N_\theta$), the local indicator is defined as
\begin{equation}
    q_{u_z}^{(k)}(t)
    =
    \frac{
        \int_{|z|<z_\mathrm{lim}} \int_{0.90R}^{0.95R}
        u_z(r,z,\theta_k,t)\, r \, \mathrm{d}r \, \mathrm{d}z
    }{
        \int_{|z|<z_\mathrm{lim}} \int_{0.90R}^{0.95R}
         \, r \, \mathrm{d}r \, \mathrm{d}z
    } \; \text{,}
    \label{eq:uz_plane_avg}
\end{equation}
where $ r\,\mathrm{d}r\,\mathrm{d}z$ is the meridional area element. 
Near the outer wall, the axial velocity is directly related to the large-scale meridional circulation: a positive (negative) $q_{u_z}$ indicates a net upward (downward) axial flux at the periphery, which discriminates between the two asymmetric recirculation states. This near-wall quantity thus provides a geometrically local and computationally inexpensive proxy for the symmetry-breaking dynamics.
The metastable states are highlighted using an \ac{EMA} of the order parameter. The \ac{EMA} is defined as
\begin{equation}
    \hat{\varphi}(t)
=
\frac{1}{T_s}
\int_{-\infty}^{t}
\exp\!\left(-\frac{t-\tau}{T_s}\right)
\varphi(\tau) \, \mathrm{d}\tau \; \text{,}
    \label{eq:varphi_EMA}
\end{equation}
for a variable $\varphi$. $T_s$ is the averaging time scale and is chosen as $T_s=5$ impeller revolutions, based on observations of the decorrelation time of the order parameter. This choice ensures that the averaging filters out fast turbulent fluctuations while preserving the slow, large-scale dynamics associated with metastable flow states.
The system is assigned to state A (respectively state B) at time $t$ if the \ac{EMA}-filtered dimensionless order parameter $\hat{q}_{u_z}^{(k)}/\Omega R$ has remained continuously above a threshold $\gamma_A = 0.1$ (respectively below $\gamma_B = -0.1$) over a trailing dwell window of duration $T_d = 5$ impeller revolutions; otherwise the state is left undefined. This dwell-time requirement prevents brief threshold crossings due to residual turbulent fluctuations from being misclassified as state transitions, so that only genuine, sustained excursions into either macroscopic configuration are labeled.
While fast turbulent fluctuations are visible at short time scales, the signal also exhibits slow variations characterized by extended plateaus separated by relatively rapid transitions. These plateaus correspond to quasi-persistent flow states, indicating that the system remains trapped in distinct large-scale configurations for durations that far exceed the characteristic turbulent time. The \ac{PDF} of $\hat{q}_{u_z}$ is bimodal (Fig.~\ref{fig:order_param_histogram}), reflecting the coexistence of two preferred macroscopic states. Such behavior is a hallmark of metastable dynamics in the von Kármán configuration and is qualitatively consistent with experimental observations \cite{delatorre2007slow,cortet2010experimental,cortet2011susceptibility}.

Both metastable states are asymmetric and consist of two recirculation cells of unequal size and intensity (Fig.~\ref{fig:order_parameter_signal_and_velocity_conditional_mean}(b)). In one case, a dominant large-scale circulation cell occupies the upper part of the vessel, while a smaller secondary cell persists near the bottom impeller. In the other case, the roles of the two cells are reversed, with the dominant circulation located below the mid-plane and a weaker cell remaining above. These two asymmetric configurations correspond to symmetry-related metastable states of the von Kármán flow. Their coexistence, together with the intermittent transitions observed in the time evolution of $\hat{q}_{u_z}^{(k)}(t)$, is indicative of spontaneous symmetry breaking of the mean flow topology. An animation of a meridional plane showing the \ac{EMA}-filtered velocity field is provided in the Supplemental Material~\cite{SupplMat} and illustrates the transition between the two metastable states.

\begin{figure}[!htbp]
    \centering
    \begin{tikzpicture}
        \node[anchor=south west, inner sep=0] (img)
            {\includegraphics[width=0.68\linewidth]{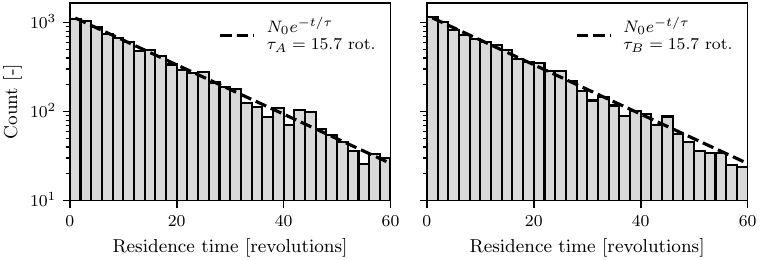}};
        \node[anchor=south west, font=\small] at (img.south west) {(a)};
        \node[anchor=south west, font=\small, xshift=3mm] at (img.south) {(b)};
    \end{tikzpicture}
    \caption{Histogram of residence times in (a) state A and (b) state B, obtained from the \ac{EMA}-filtered order parameter $\hat{q}_{u_z}^{(k)}(t)$ and pooled over all azimuthal planes $k=1,\ldots,N_\theta$. Dashed lines show exponential-decay fits to each distribution, with fitted residence times reported in the legend.}
    \label{fig:dwell_time_histogram_hat}
\end{figure}

To quantify the characteristic lifetime of each metastable state, the residence time is computed, i.e., the length of each uninterrupted interval during which the system remains classified in state A or state B, pooled over all azimuthal planes. The resulting distributions of residence times (Fig.~\ref{fig:dwell_time_histogram_hat}) decay exponentially, suggestive of a memoryless escape process, consistent with noise-driven transitions between metastable states rather than a sharp deterministic escape time. Residence-time distributions following an exponential decay law have been consistently reported in von Kármán experiments~\cite{delatorre2007slow,burguete2009hysteresis}. An exponential-decay fit to each distribution yields residence times of $\tau_A \approx 15.6$ and $\tau_B \approx 15.7$ impeller revolutions for states A and B, respectively, quantifying the persistence of the large-scale asymmetric circulation before a spontaneous transition occurs. The near-equality of $\tau_A$ and $\tau_B$ is consistent with the symmetry-related nature of the two metastable states when the two impellers counter-rotate at the same frequency.

The ability of the LES to sustain these asymmetric states and transition between them shows that the large-scale metastable dynamics characteristic of the von Kármán configuration are captured by the present modeling approach. Importantly, the transitions between states occur spontaneously, without any imposed time-dependent forcing, demonstrating that the present impeller model allows the intrinsic low-frequency dynamics of the system to emerge naturally.

\subsubsection{Azimuthal structure of the metastable states}

\begin{figure}[!htbp]
    \centering
    \includegraphics[width=0.42\linewidth]{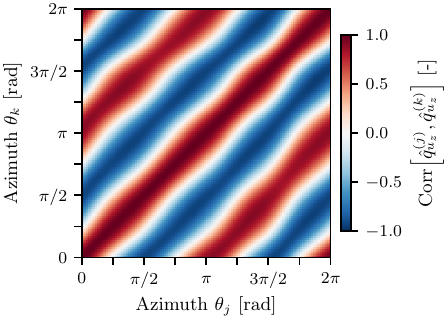}
    \caption{Equal-time Pearson correlation matrix of the \ac{EMA}-filtered order parameter $\hat{q}_{u_z}^{(k)}$ between all pairs of azimuthal planes $\theta_j,\theta_k$, $j,k=1,\ldots,N_\theta$. Diametrically opposite planes ($|\theta_j-\theta_k|=\pi$) are positively correlated. Conversely, planes separated by a quarter turn ($|\theta_j-\theta_k|=\pi/2$) are anti-correlated, indicating that the metastable states have a dominant azimuthal wavenumber $m=2$.}
    \label{fig:correlation_matrix_hat}
\end{figure}

The analysis so far has focused on a single angular position. However, the \ac{LES} resolves the full azimuthal extent of the flow, which allows the azimuthal organization of flow states to be characterized. To this end, the equal-time Pearson correlation coefficient of the \ac{EMA}-filtered order parameter $\hat{q}_{u_z}^{(k)}$ is computed between all pairs of azimuthal planes $\theta_j,\theta_k$, $j,k=1,\ldots,N_\theta$ (Fig.~\ref{fig:correlation_matrix_hat}). Diametrically opposite planes ($|\theta_j-\theta_k|=\pi$) are positively correlated, whereas planes separated by a quarter turn ($|\theta_j-\theta_k|=\pi/2$) are anti-correlated. This pattern is the signature of a dominant azimuthal wavenumber $m=2$: the metastable states identified from the meridional-plane statistics are not axisymmetric but consist of two pairs of counter-rotating recirculation cells distributed around the circumference of the vessel. 
This is in line with the proper orthogonal decomposition analysis of experimental measurements by Podvin and Dubrulle~\cite{podvin2018largescale}, who found the symmetry-breaking dynamics to be dominated by a few energetic vessel-scale modes of even azimuthal wavenumber, with a dominant $m=2$ contribution.

\begin{figure}[!htbp]
    \centering
    \includegraphics[width=1.0\linewidth]{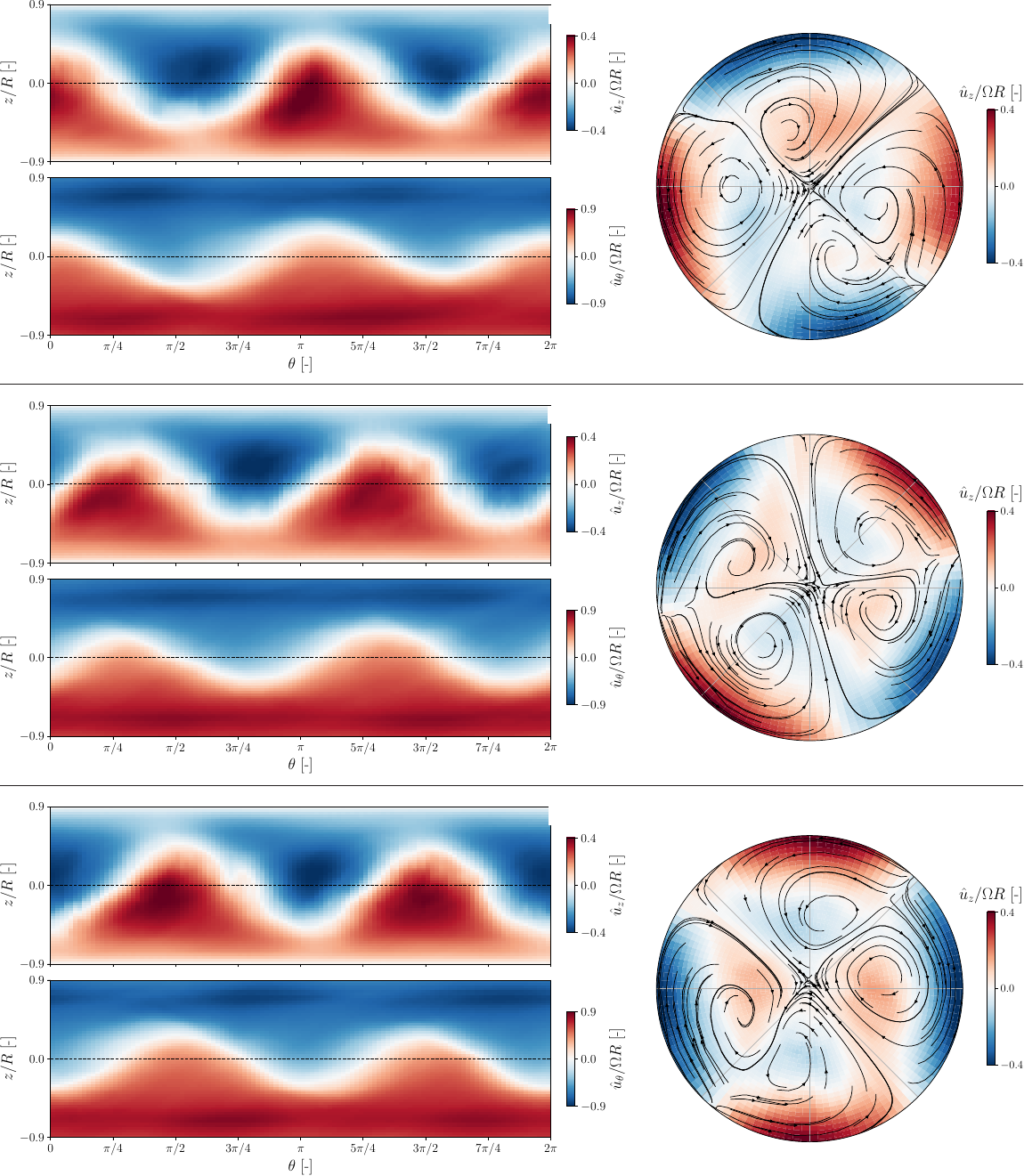}
    \caption{Snapshots of the \ac{EMA}-filtered velocity field at three distinct instants, each corresponding to a metastable state whose positive order-parameter lobe is centered at a different azimuthal position ($\theta=0$, $\pi/4$ and $\pi/2$, respectively). The \ac{EMA}-filtered velocity field is obtained by applying Eq.~(\ref{eq:varphi_EMA}) to the velocity field at each grid point. Left: unrolled cylindrical section at the second cell from the outer wall colored by \ac{EMA}-filtered dimensionless axial velocity $\hat u_z/\Omega R$ and azimuthal velocity $\hat u_\theta/\Omega R$. Right: mid-height cut colored by \ac{EMA}-filtered dimensionless axial velocity $\hat u_z/\Omega R$, with streamlines of the in-plane $(\hat u_r,\hat u_\theta)$ velocity. A four-cell pattern is clearly visible in each mid-height cut, with axial velocity positively correlated between diametrically opposite sectors ($\Delta\theta=\pi$) and anti-correlated between sectors separated by a quarter turn ($\Delta\theta=\pi/2$).}
    \label{fig:azimuth_cuts_1}
\end{figure}

\begin{figure}[!htbp]
    \centering
    \hspace*{\fill}
    \includegraphics[width=0.3\linewidth]{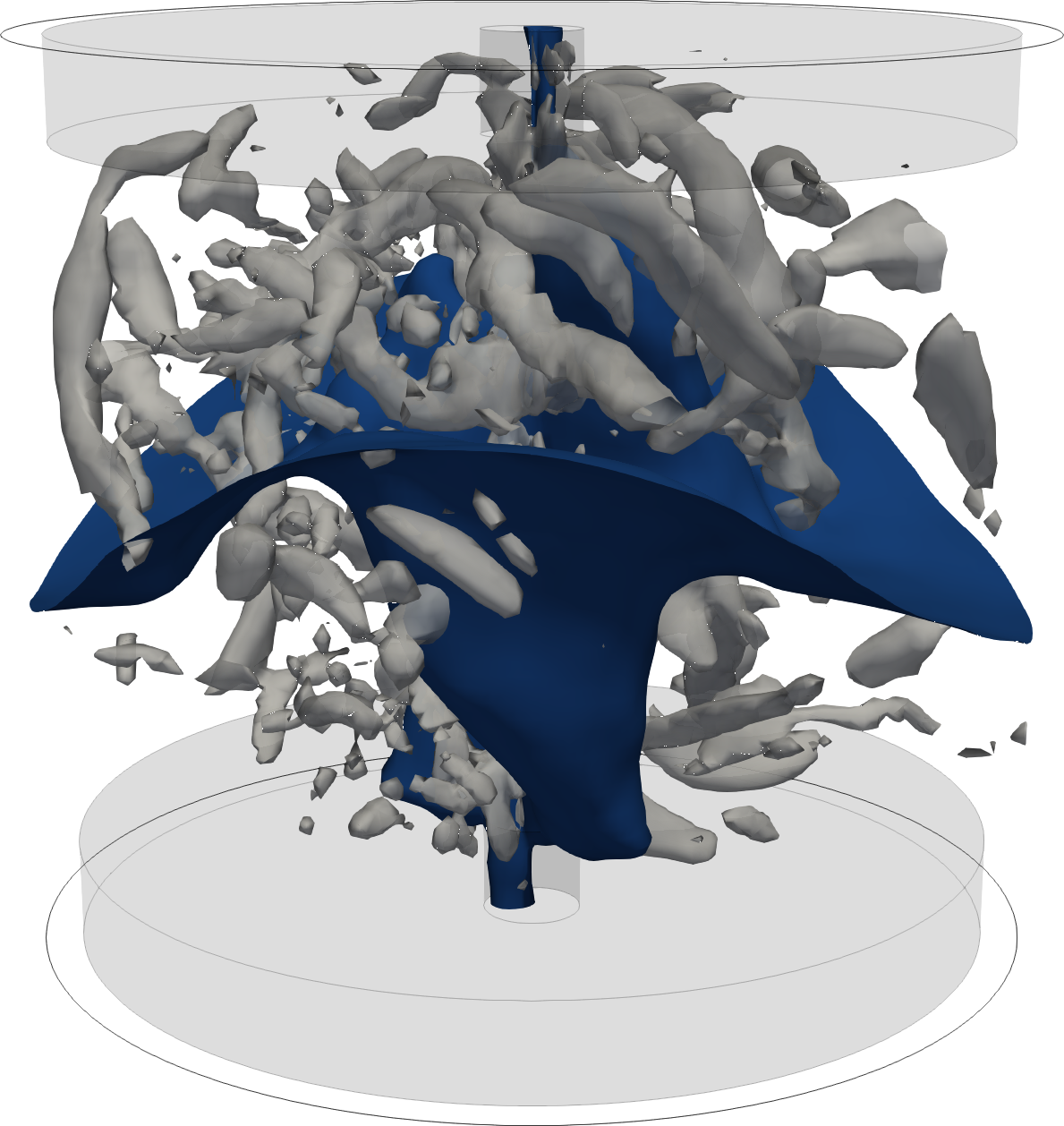}
    \hspace*{\fill}
    \includegraphics[width=0.3\linewidth]{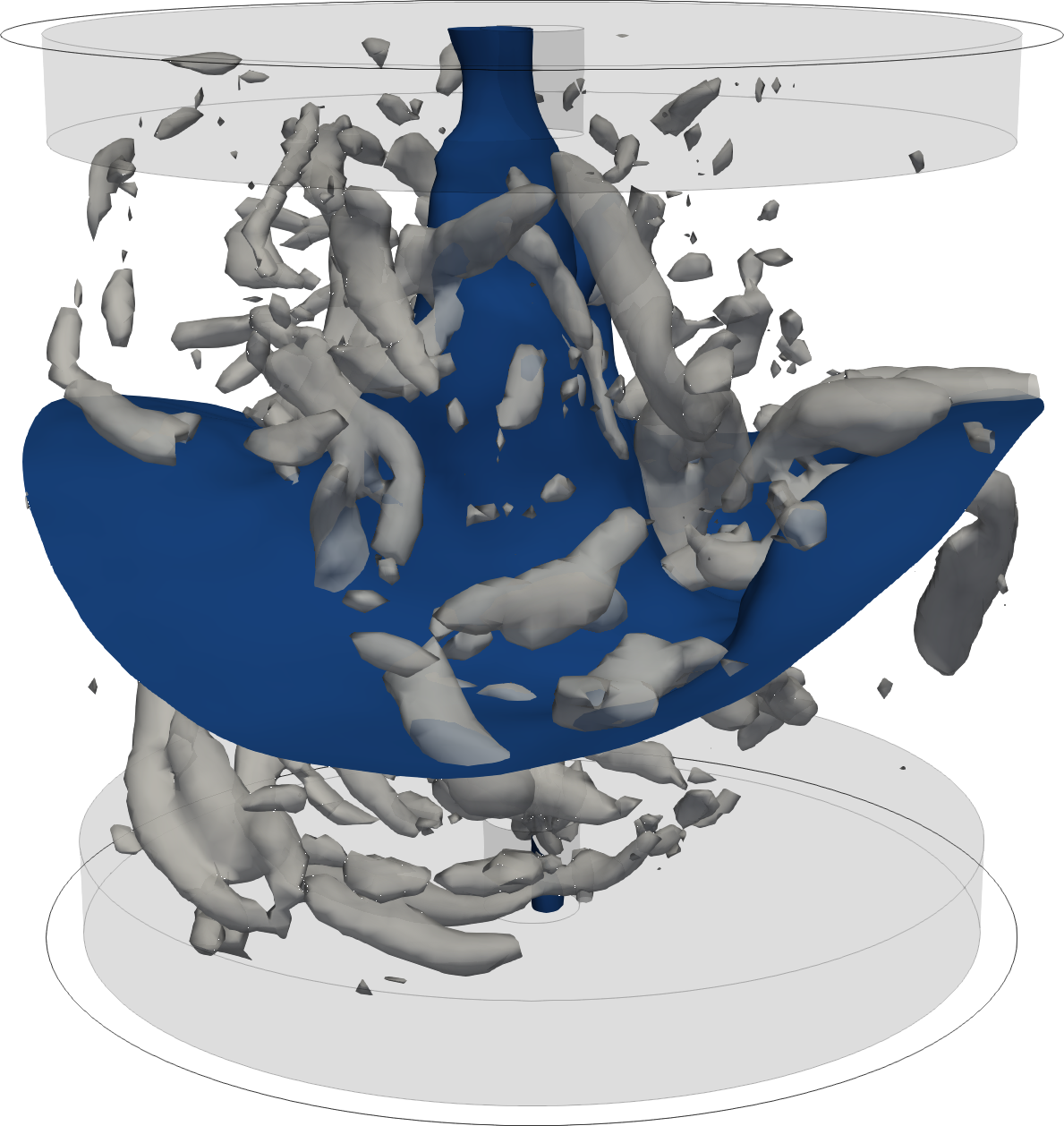}
    \hspace*{\fill}
    \caption{Three-dimensional flow visualizations at two distinct instants. Gray: iso-surfaces of the Q-criterion, $Q = \tfrac{1}{2}\left(\|\boldsymbol{\Omega}\|^2 - \|\mathbf{S}\|^2\right)$, where $\boldsymbol{\Omega}$ and $\mathbf{S}$ are the antisymmetric and symmetric parts of the velocity gradient tensor, respectively, highlighting small-scale turbulent vortical structures. Blue: zero-level iso-surface of the \ac{EMA}-filtered azimuthal velocity, $\hat u_\theta = 0$, indicating the location of the shear layer separating the two counter-rotating cells and revealing the large-scale structure associated with the slow instability. The translucent gray volumes at the top and bottom of the vessel indicate the impeller regions.}
    \label{fig:isoQcrit}
\end{figure}

This wavenumber-2 organization is confirmed directly in physical space in Fig.~\ref{fig:azimuth_cuts_1}, which shows, for three distinct instants, an unrolled cylindrical section near the outer wall together with a cross-sectional cut normal to the cylinder axis, at the mid-plane of the vessel (hereafter, the mid-height cut). In each case, the mid-height cut exhibits a clear four-cell pattern, with axial velocity of alternating sign in azimuthal sectors of extent $\pi/2$: positively correlated between diametrically opposite sectors and anti-correlated between sectors separated by a quarter turn, consistent with the correlation analysis of Fig.~\ref{fig:correlation_matrix_hat}. The three instants selected have their positive order-parameter lobe centered at three different azimuthal positions ($\theta=0$, $\pi/4$ and $\pi/2$, respectively), showing that the four-cell structure is not pinned to a fixed orientation: the metastable states are degenerate with respect to a rotation of the pattern about the cylinder axis, and the wavenumber-2 organization persists regardless of its azimuthal phase. 
This four-cell organization is similar to that of the $m=2$ states found at low Reynolds number in the exactly counter-rotating flow, both numerically~\cite{nore2003mode} and experimentally~\cite{nore2005experimental}. It also relates to the visual observations of Cortet et al.~\cite{cortet2010experimental,cortet2011susceptibility}, who reported four large-scale vortices along the shear layer for Reynolds numbers between $86\,000$ and $300\,000$, a range that includes the present operating point.
Figure~\ref{fig:isoQcrit} provides a three-dimensional view of the flow at two distinct instants. The zero iso-surface of the \ac{EMA}-filtered azimuthal velocity reveals the three-dimensional structure of the shear layer separating the counter-rotating cells, while the Q-criterion iso-surfaces highlight the small-scale turbulent vortical structures present in the flow.


\begin{figure}[!htbp]
    \centering
    \includegraphics[width=1.0\linewidth]{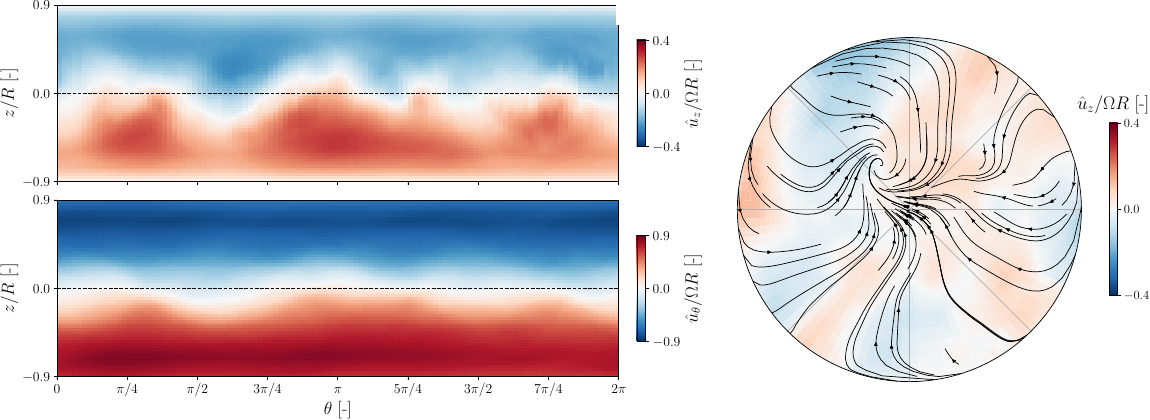}
    \caption{Snapshots of the \ac{EMA}-filtered velocity field for an instant where the state of the system is unclear. The \ac{EMA}-filtered velocity field is obtained by applying Eq.~(\ref{eq:varphi_EMA}) to the velocity field at each grid point. Left: unrolled cylindrical section at the second cell from the outer wall colored by \ac{EMA}-filtered dimensionless axial velocity $\hat u_z/\Omega R$ and azimuthal velocity $\hat u_\theta/\Omega R$. Right: mid-height cut colored by \ac{EMA}-filtered dimensionless axial velocity $\hat u_z/\Omega R$, with streamlines of the in-plane $(\hat u_r,\hat u_\theta)$ velocity. No four-cell structure is observed, reflecting the loss of the wavenumber-2 organization.}
    \label{fig:azimuth_cuts_2}
\end{figure}

The azimuthal phase of the four-cell pattern is not static: inspection of the corresponding animation in the Supplemental Material~\cite{SupplMat} shows that the structure typically drifts slowly around the circumference, rotating in either direction in an apparently random fashion, rather than remaining locked at a fixed $\theta$. Transitions between metastable states, in the sense of Fig.~\ref{fig:order_parameter_signal_and_velocity_conditional_mean}, most often proceed through this gradual azimuthal drift. In rarer instances, however, the four-cell organization instead collapses locally before reforming, without a clearly defined rotation; Fig.~\ref{fig:azimuth_cuts_2} shows an example of such an instant, for which the state of the system is ambiguous. In this case, the mid-height cut no longer exhibits two distinct pairs of counter-rotating cells but instead a single dominant recirculation, reflecting a momentary loss of the wavenumber-2 organization as the flow reorganizes from one metastable configuration to another. 
This phenomenology is reminiscent of the near-heteroclinic cycles observed at low Reynolds number~\cite{nore2003mode,nore2005experimental}, in which the system switches between $m=2$ states rotated by $\pi/2$ through rapid excursions dominated by other wavenumbers.
The observed azimuthal drift and transient reorganization describe how local state switching occurs, but do not establish the physical mechanism responsible for the emergence and evolution of the large-scale modes. These modes develop spontaneously under a steady, axisymmetric impeller target. Whether they originate from an instability of the turbulent mean flow remains to be established.

\FloatBarrier
\section{Conclusion}

A two-stage optimization strategy has been introduced to model impeller-driven von Kármán flow in \ac{LES} without resolving blade geometry.
The impeller-region velocity is expressed as a low-order B-spline parametrization whose coefficients are first determined by \ac{ODIL}, which solves an inverse problem constrained by the \ac{RANS} equations, 
experimental bulk-flow and torque measurements 
to recover a physically consistent representation of the impeller-induced flow without requiring direct access to the impeller-region velocities.
The inferred velocities constitute a compact parametrization of the impeller forcing that is then coupled to the \ac{LES} through momentum source terms.
The coefficients are subsequently optimized by \ac{CMA-ES} coupled directly to the \ac{LES} solver, minimizing the mismatch between LES and experimental time-averaged velocity and torque.
This \ac{LES}-coupled optimization corrects the modeling error of the \ac{RANS}-based inference and yields an improved mean-flow prediction.
Crucially, the resulting LES captures metastable large-scale flow states and spontaneous transitions without imposing time-dependent forcing, enabling intrinsic instabilities characterizing the slow-regime behavior of the system to be observed. The metastable states are further shown to be spatially structured, consisting of an alternating four-cell pattern around the vessel circumference that slowly drifts in azimuth, with local state transitions typically proceeding through this drift. The simulation of such flow dynamics is inaccessible to blade-resolved approaches at comparable Reynolds numbers due to their prohibitive computational cost. The proposed approach therefore provides a flexible and computationally efficient framework for studying slow, large-scale dynamics in impeller-driven turbulent flows, without explicitly resolving the blade geometry. 
The impeller parametrization has been assessed only for the configuration for which it was optimized. Its applicability to other vessel aspect ratios, blade geometries, and Reynolds numbers remains untested.
Future work will investigate this transferability and develop more generalizable impeller parametrizations (e.g. to account for Reynolds number effects) if needed.
In addition, the calibration of the forcing coefficient $\alpha$ of the impeller model (Eq.~(\ref{eq:forcing_LES})) will be investigated further. Indeed, the relaxation time scale of the forcing affects the distribution of the system order parameter (Appendix~\ref{appendix:alpha_sensitivity}). This motivates including $C_\alpha$, in $\alpha = C_\alpha |\Omega|$, among the impeller model parameters to be optimized (Appendix~\ref{appendix:alpha_optimization}) and incorporating experimental instability statistics, once available, into the optimization objective function.
Further work will focus on identifying the physical origin of the large-scale modes and characterizing their statistics over a wider range of operating conditions. Analysis of the azimuthal mode amplitudes and phases, together with a stability analysis of the turbulent mean flow, could help clarify the mechanisms driving their emergence, drift, and transient reorganization.

\begin{acknowledgments}
This research was primarily supported by the Defense Advanced Research Projects Agency (DARPA) under the Automated Prediction Aided by Quantized Simulators (APAQuS) program, Grant No. HR00112490529.
The authors gratefully acknowledge Bérengère Dubrulle for sharing the experimental data used in this study.
The authors also thank Adam Pua and Mirko Gamba for helpful discussions.
\end{acknowledgments}

\section*{Data availability}

The experimental \ac{PIV} and torque data that support the findings of this study are available in Ref.~\cite{cappanera_turbulence_2021}.

\appendix

\FloatBarrier
\section{Reference experimental data}
\label{appendix:experiments}

\subsection{Experimental apparatus}

The reference \ac{PIV} velocity fields used throughout this work were acquired by Cappanera et al.~\cite{cappanera_turbulence_2021} on the von~Kármán flow device described therein. The setup consists of a cylindrical vessel filled with water and driven by two counter-rotating impellers of TM87 geometry~\cite{Ravelet2005}. The dimensions of the von~Kármán flow device are given in Table~\ref{tab:geometry}. The impellers feature $8$ curved blades with a blade angle of $72^\circ$. The reference data used in this work correspond to the operating point $\mathrm{Re}=300,000$, with the impellers counter-rotating at the same absolute frequency (i.e., $\Omega_1=-\Omega_2$), both spinning with the concave side of the blades leading. Velocity fields in the meridional plane are measured by \acf{SPIV}, and the torque applied to each impeller shaft is independently measured using torque meters on the driving motors. We refer to Cappanera et al.~\cite{cappanera_turbulence_2021} for further details on the experimental apparatus and measurement procedures.

\subsection{Data regularization}

Before being used as reference data, the mean velocity field undergoes two regularization steps: enforcement of physical symmetry and axis conditions, and a divergence-free projection. Each step is described below.

\subsubsection{Midplane and axis symmetry}

The von~Kármán configuration with equal impeller speeds is statistically symmetric about the midplane $z = 0$, and the cylindrical mean velocity components obey definite parities:
\begin{equation}
  \bar{u}_r(r, z) = \bar{u}_r(r, -z), \qquad
  \bar{u}_z(r, z) = -\bar{u}_z(r, -z), \qquad
  \bar{u}_\theta(r, z) = -\bar{u}_\theta(r, -z).
\end{equation}
These parities are enforced by replacing each field with its symmetrized version,
\begin{equation}
    \bar{u}_r \leftarrow \tfrac{1}{2}(\bar{u}_r + \bar{u}_r^{-}), \quad
    \bar{u}_z \leftarrow \tfrac{1}{2}(\bar{u}_z - \bar{u}_z^{-}), \quad
    \bar{u}_\theta \leftarrow \tfrac{1}{2}(\bar{u}_\theta - \bar{u}_\theta^{-}),
\end{equation}
where the superscript $^{-}$ denotes the field evaluated at $-z$. At the midplane row ($z = 0$) the odd components are additionally set to zero exactly: $\bar{u}_z = \bar{u}_\theta = 0$.

On the symmetry axis ($r = 0$), regularity of an axisymmetric field requires $\bar{u}_r = 0$, $\bar{u}_\theta = 0$, and zero radial gradient of $\bar{u}_z$; all three conditions are enforced explicitly.

\subsubsection{Divergence-free projection}

Experimental PIV data are not exactly solenoidal due to measurement noise and interpolation errors.
A weighted least-squares divergence correction is applied to the pair $(\bar{u}_r, \bar{u}_z)$ to minimize the cylindrical continuity residual
\begin{equation}
  \mathcal{D}(\bar{u}_r, \bar{u}_z) = \frac{1}{r}\frac{\partial (r\, \bar{u}_r)}{\partial r} + \frac{\partial \bar{u}_z}{\partial z}.
\end{equation}
The discrete divergence is evaluated with standard second-order finite differences on the uniform $(r, z)$ \ac{SPIV} grid. A least-squares correction $\mathbf{s} = (s_r, s_z)$ is then computed by solving
\begin{equation}
  \min_{\mathbf{s}}\; \|\mathbf{s}\|_2^2 \quad \text{subject to} \quad \mathbf{D}\,(\bar{\mathbf{u}} - \mathbf{s}) \approx \mathbf{0},
\end{equation}
where $\mathbf{D}$ is the discrete divergence operator and $\bar{\mathbf{u}} = (\bar{u}_r, \bar{u}_z)$ is the measured field. The system is solved with LSQR.
The final corrected field is used as reference data throughout this work. A one-cell-wide boundary strip along all four edges of the domain is removed before comparison, as boundary cells carry one-sided stencil residuals that are not reduced by the interior projection.

\FloatBarrier
\section{Gauss--Newton algorithm for ODIL inference}
\label{appendix:gn_odil}

The ODIL inference problem is posed as the minimization of a nonlinear least-squares objective built from a stacked residual vector $\mathbf r(\mathbf v)$, which collects the discretized governing-equation residuals together with data-misfit terms (and any additional constraints). Here $\mathbf v$ denotes the optimization state, obtained by concatenating all unknown degrees of freedom (flow variables on the numerical grid and the spline coefficients used to parametrize the impeller-region velocities). We solve
\begin{equation}
    \min_{\mathbf v}\ \|\mathbf r(\mathbf v)\|_2^2 \; \text{.}
\end{equation}
At iteration $k$, Gauss--Newton linearizes $\mathbf r(\mathbf v)$ about $\mathbf v^k$ and computes an update $\Delta \mathbf v$ by solving the system $(\mathbf J^\top \mathbf J + \lambda \mathbf I)\Delta\mathbf v = -\mathbf J^\top \mathbf r$, where $\mathbf J(\mathbf v^k)=\partial \mathbf r/\partial \mathbf v$ and $\lambda\ge 0$ is a small damping parameter ($\lambda=10^{-8}$ in this work). The linear system is solved using GMRES by applying the operator $\mathcal A(\mathbf w)=\mathbf J^\top(\mathbf J\mathbf w)+\lambda \mathbf w$ through Jacobian--vector and transposed-Jacobian--vector products computed with automatic differentiation, without forming $\mathbf J$ explicitly. The step is then accepted using a backtracking line search enforcing an Armijo decrease condition on $\|\mathbf r\|_2$. Algorithm~\ref{alg:gauss_newton_vktts} summarizes the procedure.

\begin{figure}[!htbp]
\centering
\begin{minipage}{1.0\linewidth}
\begin{algorithmic}[1]
\Require Initial state vector $\mathbf{v}^0$ (fields + spline coefficients), max iterations $K$, Armijo parameter $c_1$, max line-search steps $N_{\mathrm{ls}}$, damping $\lambda\ge 0$
\For{$k=0,\dots,K-1$}
  \State $\mathbf r \gets \mathbf r(\mathbf{v}^k)$ \Comment{assemble weighted residual vector (PDE + data)}

  \State $\mathbf g \gets \mathbf J(\mathbf{v}^k)^\top \mathbf r$
  \State Define operator $\mathcal{A}(\mathbf w) = \mathbf J(\mathbf{v}^k)^\top(\mathbf J(\mathbf{v}^k) \mathbf w) + \lambda \mathbf w$
  \State Solve approximately $\mathcal{A}(\Delta \mathbf v) = - \mathbf g$ using GMRES \Comment{normal equations}

  \State $s \gets 1$
  \For{$\ell=1,\dots,N_{\mathrm{ls}}$}
    \State $\mathbf v^{\mathrm{trial}} \gets \mathbf{v}^k + s \, \Delta \mathbf v$
    \State $\mathbf r^{\mathrm{trial}} \gets \mathbf r(\mathbf v^{\mathrm{trial}})$
    \If{$\|\mathbf{r}^{\mathrm{trial}}\|_2 \le (1-c_1 s) \, \|\mathbf r\|_2$} \Comment{Armijo condition on $\|\mathbf r\|_2$}
      \State \textbf{break}
    \EndIf
    \State $s \gets s/2$
  \EndFor
  \If{$\ell > N_{\mathrm{ls}}$}
    \State $\mathbf v^{k+1} \gets \mathbf{v}^k + 10^{-2}\Delta \mathbf v$ \Comment{fallback if no sufficient decrease}
  \Else
    \State $\mathbf v^{k+1} \gets \mathbf v^{\mathrm{trial}}$
  \EndIf

\EndFor
\State \Return $\mathbf{v}^K$
\end{algorithmic}
\end{minipage}
\algorithmcaption{Gauss--Newton for ODIL residual minimization}{alg:gauss_newton_vktts}
\end{figure}

\FloatBarrier
\section{Equilibrium wall model}
\label{appendix:wall_model}

The wall model at the top, bottom and outer cylindrical walls of the vessel is derived assuming that the near-wall boundary layer is in local equilibrium. The inner part of the turbulent boundary layer is replaced by the solution to a one-dimensional equilibrium boundary-layer problem forced by the resolved LES velocity at the first off-wall cell. Let $y$ denote the wall-normal distance and $u(y)$ the wall-tangential velocity component in the inner layer. The mean velocity is assumed to obey the equilibrium momentum balance
\begin{equation}
    \frac{\mathrm d}{\mathrm d y}\left[(\nu + \nu_T)\frac{\mathrm d u}{\mathrm d y}\right] = 0 \, \text{,}
    \label{eq:wall_eq_mom_balance}
\end{equation}
where $\nu$ is the kinematic viscosity and $\nu_T(y)$ the turbulent viscosity. The wall stress reads
\begin{equation}
    \tau_w = \rho u_\tau^2 = \rho \nu \left.\frac{\mathrm d u}{\mathrm d y}\right|_{y=0} \, \text{,}
\end{equation}
where $u_\tau$ is the friction velocity. Integrating Eq.~(\ref{eq:wall_eq_mom_balance}) gives
\begin{equation}
    (\nu+\nu_T) \frac{\mathrm d u}{\mathrm d y} = u_\tau^2 \, \text{.}
\end{equation}
The turbulent eddy viscosity is obtained from a mixing-length model with a van Driest damping function
\begin{equation}
    \nu_T(y) = \left[D(y)\,\kappa y\right]^2 \left|\frac{\mathrm d u}{\mathrm d y}\right| \quad \text{with} \quad
    D(y) = 1 - \exp(-y^+/A^+) \, \text{,}
    \label{eq:ml_visc}
\end{equation}
and with $\kappa=0.4$, $A^+=25$ and $y^+ = y u_\tau/\nu$. This yields a first-order nonlinear \ac{ODE} for $u(y)$
\begin{equation}
    \nu \frac{\mathrm d u}{\mathrm d y} + \left[D(y)\,\kappa y\right]^2 \left|\frac{\mathrm d u}{\mathrm d y}\right| \frac{\mathrm d u}{\mathrm d y} = u_\tau^2 \, \text{,}
    \label{eq:ODE_eq_wall_model}
\end{equation}
subject to the boundary conditions $u(0)=0$ and $u(\Delta_y) = U_{\mathrm{LES},||}$, where $U_{\mathrm{LES},||}$ is the magnitude of the wall-parallel LES velocity at the cell adjacent to the wall and $\Delta_y$ is the wall-normal distance from the wall to the center of that cell. Equation~(\ref{eq:ODE_eq_wall_model}) with the two boundary conditions defines a one-parameter problem whose unknown is the friction velocity $u_\tau$.

The traditional approach uses friction velocity $u_\tau$ as velocity scale for non-dimensionalization, but since $u_\tau$ is the unknown quantity and to avoid iterations, one must use the only available dimensional  quantities, namely $\Delta_y$ and viscosity $\nu$ \cite{Meneveau01112020}. Working in the dimensionless form $\hat y = y/\Delta_y$, $\hat u = u\,\Delta_y/\nu$, $\hat\nu_T = \nu_T/\nu$, Eq.~(\ref{eq:ODE_eq_wall_model}) becomes
\begin{equation}
    \frac{\mathrm d \hat u}{\mathrm d \hat y} + \left[D(\hat y)\,\kappa \hat y\right]^2 \left(\frac{\mathrm d \hat u}{\mathrm d \hat y}\right)^2 = \hat u_\tau^2 \qquad \text{with} \qquad
    \hat u_\tau = u_\tau \Delta_y/\nu \, \text{,}
\end{equation}
and with boundary condition $\hat u(1) = \hat U_{\mathrm{LES},||} = U_{\mathrm{LES},||}\Delta_y/\nu = \mathrm{Re}_\Delta$. Solving for $\mathrm d\hat u/\mathrm d\hat y$ yields
\begin{equation}
    \frac{\mathrm d \hat u}{\mathrm d \hat y} = \frac{1}{2\left[D(\hat y)\,\kappa \hat y\right]^2}\left(-1 + \sqrt{1 + 4\left[D(\hat y)\,\kappa \hat y\right]^2 \hat u_\tau^2}\right) \, \text{,}
    \label{eq:eq_grad}
\end{equation}
so that the friction velocity is obtained as the root of
\begin{equation}
    \int_0^1 \frac{1}{2\left[D(\hat y)\,\kappa \hat y\right]^2}\left(-1 + \sqrt{1 + 4\left[D(\hat y)\,\kappa \hat y\right]^2 \hat u_\tau^2}\right) \mathrm d \hat y = \hat U_{\mathrm{LES},||} \, \text{,}
\end{equation}
for given $\hat U_{\mathrm{LES},||}$. In practice, $u_\tau(\mathrm{Re}_\Delta)$ is evaluated through the $\mathrm{Re}_\tau$--$\mathrm{Re}_\Delta$ fit of Meneveau~\cite{Meneveau01112020}. The wall stress imposed on the LES is then
\begin{equation}
    \tau_w = \rho u_\tau^2 \hat{\mathbf t} \, \text{,}
\end{equation}
with $\hat{\mathbf t}$ the unit vector aligned with the wall-parallel LES velocity at the matching cell.

Numerically, the wall stress is not imposed explicitly; it is recovered through the solver's viscous flux at the wall face using a ghost-cell method. The tangential velocity of the wall-adjacent ghost cell is extrapolated from the interior using the equilibrium gradient from Eq.~(\ref{eq:eq_grad}), so that the discrete gradient at the first off-wall cell matches that of the equilibrium solution. The eddy viscosity of the first off-wall cell is set to the mixing-length value of Eq.~(\ref{eq:ml_visc}), while that of the ghost cell is chosen so that the discrete viscous flux across the wall face equals $\rho u_\tau^2 \hat{\mathbf t}$. The wall model therefore enters the LES solely through the ghost velocity and the near-wall eddy viscosity, leaving the interior scheme untouched.

\FloatBarrier
\section{Grid sensitivity analysis}\label{appendix:grid_sensitivity}

The statistics of the slow variations of the order parameter require long simulation durations, making their evaluation on grids finer than the reference grid computationally prohibitive within the present study. Grid sensitivity is therefore assessed by progressively coarsening the reference grid from $N_r \times N_\theta \times N_z = 35 \times 112 \times 62$ to $26 \times 84 \times 46$ and $17 \times 56 \times 30$. The comparison considers both the mean velocity field and the \ac{PDF} of the \ac{EMA}-filtered order parameter.

All three simulations use the same \ac{ODIL}-inferred B-spline coefficients $\mathbf{c}^{(0)}$, without further adjustment to the grid. The \ac{CMA-ES} refinement is not applied in this analysis because it optimizes the impeller parametrization for a given grid. Repeating this optimization on each grid could compensate for resolution-dependent changes in the mean flow and thereby mask grid sensitivity. Keeping the inferred coefficients fixed allows the effect of grid resolution to be assessed for the same impeller parametrization.

\begin{figure}[!htbp]
    \centering
    \includegraphics[width=0.7\linewidth]{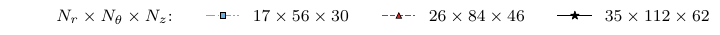}
    \par\vspace{0.5em}
    \begin{tikzpicture}
        \node[anchor=south west, inner sep=0] (img)
            {\includegraphics[width=0.325\linewidth]{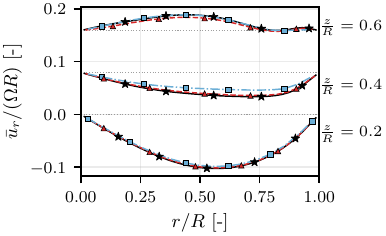}};
        \node[anchor=south west, font=\small] at (img.south west) {(a)};
    \end{tikzpicture}
    \hfill
    \begin{tikzpicture}
        \node[anchor=south west, inner sep=0] (img)
            {\includegraphics[width=0.325\linewidth]{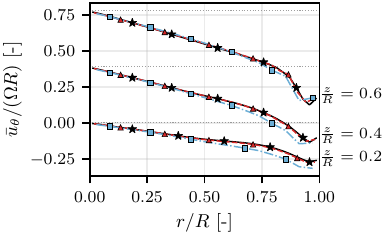}};
        \node[anchor=south west, font=\small] at (img.south west) {(b)};
    \end{tikzpicture}
    \hfill
    \begin{tikzpicture}
        \node[anchor=south west, inner sep=0] (img)
            {\includegraphics[width=0.325\linewidth]{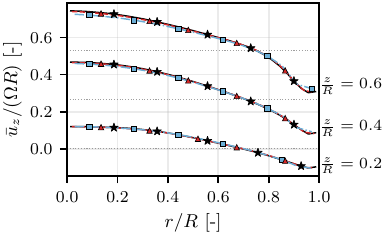}};
        \node[anchor=south west, font=\small] at (img.south west) {(c)};
    \end{tikzpicture}
    \caption{Radial profiles of the mean velocity components (a) $\bar{u}_r/(\Omega R)$, (b) $\bar{u}_\theta/(\Omega R)$, and (c) $\bar{u}_z/(\Omega R)$ at three axial positions $z/R$, for three grid resolutions. Profiles are offset by the local zero-velocity baseline for clarity (cascade representation).}
    \label{fig:grid_sensitivity_radial_profiles}
\end{figure}

\begin{figure}[!htbp]
    \centering
    \includegraphics[width=0.45\linewidth]{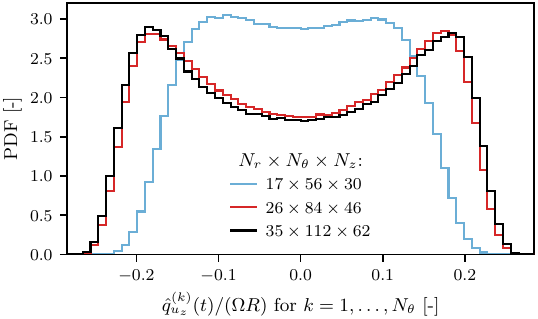}
    \caption{\ac{PDF} of the \ac{EMA}-filtered dimensionless order parameter $\hat{q}_{u_z}^{(k)}(t)/(\Omega R)$ for three grid resolutions, assembled by sampling at all azimuthal planes $k=1,\ldots,N_\theta$ for each grid.}
    \label{fig:grid_sensitivity_order_param_histogram}
\end{figure}

The radial profiles of the mean velocity components (Fig.~\ref{fig:grid_sensitivity_radial_profiles}) show little sensitivity to grid resolution over the range considered. The only noticeable difference occurs in the azimuthal velocity near the outer cylindrical wall, $r=R$, on the coarsest grid. The order-parameter distributions (Fig.~\ref{fig:grid_sensitivity_order_param_histogram}) are also nearly unchanged between the reference and intermediate grids, with both retaining a bimodal \ac{PDF}. On the coarsest grid, however, the bimodal distribution is lost and the flow instability vanishes. This shows that agreement of the mean velocity profiles alone is insufficient to assess whether the grid captures the large-scale metastable dynamics.

These results support the use of the $35 \times 112 \times 62$ grid throughout the study: both the mean flow and the order-parameter distribution remain insensitive to moderate coarsening to $26 \times 84 \times 46$, whereas further coarsening alters the flow instability.

\FloatBarrier
\section{Forcing coefficient sensitivity analysis}\label{appendix:alpha_sensitivity}

\begin{figure}[!htbp]
    \centering
    \includegraphics[width=0.75\linewidth]{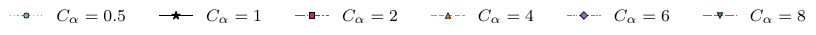}
    \par\vspace{0.5em}
    \begin{tikzpicture}
        \node[anchor=south west, inner sep=0] (img)
            {\includegraphics[width=0.325\linewidth]{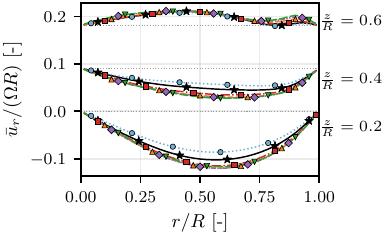}};
        \node[anchor=south west, font=\small] at (img.south west) {(a)};
    \end{tikzpicture}
    \hfill
    \begin{tikzpicture}
        \node[anchor=south west, inner sep=0] (img)
            {\includegraphics[width=0.325\linewidth]{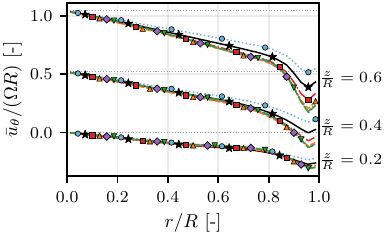}};
        \node[anchor=south west, font=\small] at (img.south west) {(b)};
    \end{tikzpicture}
    \hfill
    \begin{tikzpicture}
        \node[anchor=south west, inner sep=0] (img)
            {\includegraphics[width=0.325\linewidth]{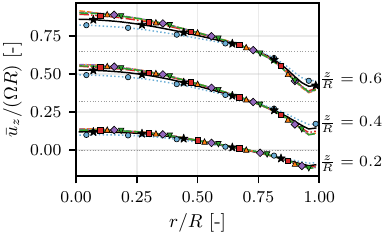}};
        \node[anchor=south west, font=\small] at (img.south west) {(c)};
    \end{tikzpicture}
    \caption{Radial profiles of the mean velocity components (a) $\bar{u}_r/(\Omega R)$, (b) $\bar{u}_\theta/(\Omega R)$, and (c) $\bar{u}_z/(\Omega R)$ at three axial positions $z/R$, for six $C_\alpha$ values. Profiles are offset by the local zero-velocity baseline for clarity (cascade representation).}
    \label{fig:alpha_sensitivity_radial_profiles}
\end{figure}

\begin{figure}[!htbp]
    \centering
    \includegraphics[width=0.45\linewidth]{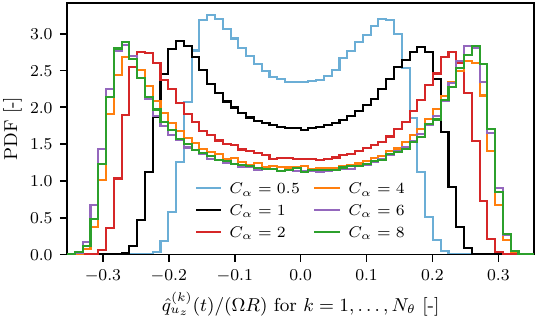}
    \caption{\ac{PDF} of the \ac{EMA}-filtered dimensionless order parameter $\hat{q}_{u_z}^{(k)}(t)/(\Omega R)$ for six $C_\alpha$ values, assembled by sampling at all azimuthal planes $k=1,\ldots,N_\theta$.}
    \label{fig:alpha_sensitivity_order_param_histogram}
\end{figure}

The influence of the forcing coefficient $\alpha$ (Eq.~(\ref{eq:forcing_LES})) is assessed by varying $C_\alpha$ in $\alpha = C_\alpha |\Omega|$ over the values $0.5$, $1$, $2$, $4$, $6$, and $8$. All simulations use the same target impeller-region velocity $\bar{\mathbf{u}}^{\mathrm{imp}}$ defined by the \ac{ODIL}-inferred coefficients $\mathbf{c}^{(0)}$, held fixed as $C_\alpha$ is varied.
The mean velocity profiles (Fig.~\ref{fig:alpha_sensitivity_radial_profiles}) show sensitivity to changes in $C_\alpha$ for $C_\alpha < 4$. When the coefficient $C_\alpha$ increases, the relaxation timescale $1/(C_\alpha |\Omega|)$ decreases, imposing the target impeller-region velocity $\bar{\mathbf{u}}^{\mathrm{imp}}$ more strongly.
For $C_\alpha \geq 4$, this has limited influence on the mean flow where the profiles nearly overlap. This indicates that the calibration of the target impeller-region velocity to achieve a certain mean flow is tied to the chosen forcing coefficient for $C_\alpha < 4$.
The order-parameter distributions (Fig.~\ref{fig:alpha_sensitivity_order_param_histogram}) also depend on $C_\alpha$. As the coefficient increases, the two peaks of the bimodal distribution move toward larger absolute values of the order parameter, while the probability density near zero decreases. The distributions become nearly unchanged for $C_\alpha \geq 4$. Bimodality is retained for all values considered, showing that this feature of the large-scale instability is robust to the tested variations in forcing strength. Its quantitative statistics, however, remain dependent on the chosen coefficient for $C_\alpha < 4$.

The value $C_\alpha=1$ used in the results presented within the main body of the manuscript is a modeling choice rather than a value calibrated against experimental instability statistics. The results therefore support a qualitative investigation of the instability, without establishing quantitative agreement with its experimental statistics. If such measurements become available, $C_\alpha$ could be included among the impeller model parameters to be optimized (Appendix~\ref{appendix:alpha_optimization}), with the measured instability statistics incorporated into the optimization objective function.

\FloatBarrier
\section{Joint optimization of the forcing coefficient and target impeller-region velocity parametrization}\label{appendix:alpha_optimization}

\begin{figure}[!htbp]
    \centering
    \includegraphics[width=0.45\linewidth]{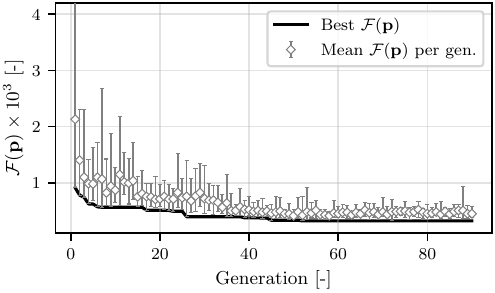}
    \caption{Evolution of the CMA-ES objective function $\mathcal{F}(\mathbf{p})$ (Eq.~(\ref{eq:objective_cmaes})) over optimization generations. The solid line shows the best objective value found up to each generation (overall best). Diamond markers show the mean $\mathcal{F}(\mathbf{p})$ within each generation, with error bars spanning the minimum and maximum values.}
    \label{fig:alpha_optim_cmaes_quality_vs_batch}
\end{figure}

To assess the choice of $C_\alpha=1$, we perform an additional \ac{CMA-ES} optimization in which $C_\alpha$ is varied jointly with the B-spline coefficients of the target impeller-region velocity parametrization. The optimization vector is augmented from the $12$ spline coefficients to $\mathbf{p} = (\mathbf{c}, C_\alpha) \in \mathbb{R}^{13}$, and the initial distribution mean is set to $\mathbf{p}^{(0)} = (\mathbf{c}^{(0)}, 1)$, where $\mathbf{c}^{(0)}$ denotes the \ac{ODIL}-inferred coefficients (Sec.~\ref{sec:odil_results}). The objective retains the mean-velocity and torque discrepancies and the physical penalty of Eq.~(\ref{eq:objective_cmaes}), with the \ac{LES} predictions now depending on both $\mathbf{c}$ and $C_\alpha$. The objective function stabilizes after approximately $60$ generations (Fig.~\ref{fig:alpha_optim_cmaes_quality_vs_batch}), indicating convergence on a similar generation count to the fixed-$C_\alpha$ optimization reported in Sec.~\ref{sec:cmaes_results}.

Table~\ref{tab:alpha_optim_mse} compares the mean-velocity and torque errors of the best individual found for the original and joint parametrizations. Both yield similar \ac{MSE} values and therefore comparable agreement with the experimental measurements. The joint optimization converges to $C_\alpha^\star = 1.13$, close to the fixed value $C_\alpha=1$ used in Sec.~\ref{sec:cmaes_results}. These results support the choice of $C_\alpha = 1$ for optimization against mean velocity and torque.
However, this does not establish whether this $C_\alpha$ value reproduces the experimental instability statistics. As shown in Appendix~\ref{appendix:alpha_sensitivity}, $C_\alpha$ affects the order-parameter distribution, which is not constrained by the present objective function. Calibrating the forcing coefficient against these statistics will require their experimental characterization and inclusion in the optimization objective, and is left for future work.

\begin{table}[!htbp]
    \centering
    \caption{\ac{MSE} relative to the experimental data for the best individual found over the \ac{CMA-ES} optimization. The velocity-field \ac{MSE} is $N_p^{-1}\|\langle\tilde{\mathbf{u}}\rangle_T-\bar{\mathbf{u}}^e\|_2^2$, and the torque \ac{MSE} is $(\langle Q\rangle_T-\bar{Q}^e)^2$. These are the unweighted discrepancy terms in Eq.~(\ref{eq:objective_cmaes}).}
    \label{tab:alpha_optim_mse}
    \begingroup
    
    \begin{ruledtabular}
    \begin{tabular}{lccc}
    Parametrization & Number of parameters & Velocity-field MSE & Torque MSE \\
    \colrule
    Original ($C_\alpha=1$ fixed) & $12$ & $1.133 \times 10^{-4}$ & $2.959 \times 10^{-4}$ \\
    Joint ($C_\alpha$ optimized) & $13$ & $1.235 \times 10^{-4}$ & $2.895 \times 10^{-4}$ \\
    \end{tabular}
    \end{ruledtabular}
    
    \endgroup
\end{table}

\FloatBarrier
\bibliography{sample}

\end{document}